%
%

\input amstex
\documentstyle{amsppt}
\magnification=1200
\pagewidth{15.2truecm}
\pageheight{22truecm}
\TagsOnRight
\CenteredTagsOnSplits

\define \<{\langle}
\define \>{\rangle}
\define \ce{Cend$_1$}

\define \p{\partial}
\define \cpx{\Bbb C[\partial, x]}
\define \cp{\Bbb C[\partial]}
\define \la{\lambda}
\define \s{\sigma}
\define \ep{\epsilon}
\define \CC{\Bbb C}

\define \ZZ{\Bbb Z}

\define \ch {\hbox{ch} }

\define \Alg{ \hbox{Alg}}
\define \Cend{ \hbox{Cend}}

\define \Diff{ \hbox{Diff}}
\define \Mat{ \hbox{Mat}}



\topmatter
\title  On the classification of subalgebras of Cend$_N$ and $gc_N$
\endtitle

\rightheadtext{On the classification of subalgebras of Cend$_N$ and $gc_N$}

\leftheadtext{C. Boyallian, V. G. Kac and  J. I. Liberati}

\author{ Carina Boyallian, Victor G. Kac and Jose I. Liberati} \endauthor

\address Ciem - FAMAF Universidad Nacional de C\'ordoba - (5000) C\'ordoba,
Argentina \endaddress
\email boyallia\@mate.uncor.edu \endemail

\address  Department of Mathematics, MIT, Cambridge, MA 02139, USA
\endaddress
\email kac\@math.mit.edu  \endemail

\address Ciem - FAMAF Universidad Nacional de C\'ordoba - (5000) C\'ordoba,
Argentina \endaddress
\email liberati\@mate.uncor.edu \endemail



\abstract
The problem of classification of infinite subalgebras of Cend$_N$ and of $gc_N$
that  acts irreducibly on $\Bbb C[\partial]^N$ is discussed in this paper.
\endabstract
\endtopmatter
\document


\head 0. Introduction\endhead

\vskip .3cm

Since the pioneering papers \cite{BPZ} and \cite{Bo}, there has been a great
deal of work towards understanding of the algebraic structure underlying the
notion of the operator product expansion (OPE) of chiral fields of a
conformal field theory.  The singular part of the OPE encodes the commutation
relations of fields, which leads to the notion of a Lie conformal algebra
\cite{K1-2}.

In the past few years a structure theory \cite{DK}, representation theory
\cite{CK, CKW} and cohomology theory \cite{BKV} of finite Lie conformal algebras
has been developed.

The associative conformal algebra Cend$_N$ and the corresponding general Lie
conformal algebra $gc_N$ are the most important examples of simple conformal
algebras which are not finite (see Sect. 2.10 in \cite{K1}). One of the
most urgent open problems of the theory of conformal algebras is the
classification of infinite   subalgebras of Cend$_N$ and of $gc_N$ which
act irreducibly on $\cp^N$. (For a classification of such finite algebras, in
 the associative case see
Theorem~5.2 of the present paper, and in the (more difficult) Lie
case see \cite{CK} and \cite{DK}.)

The classical Burnside theorem states that any subalgebra of the
matrix algebra $\Mat_N \CC$ that acts irreducibly on $\CC^N$ is
the whole algebra $\Mat_N \CC$.  This is certainly not true
for  subalgebras of $\Cend_N$ (which is the ``conformal''
analogue of $\Mat_N \CC$).  There is a family of infinite
subalgebras $\Cend_{N,P}$ of $\Cend_N$, where $P(x) \in \Mat_N
\CC [x]$, $\det P (x) \neq 0$, that still act irreducibly on $\CC
[\partial]^N$.  One of the conjectures of \cite{K2} states that
there are no other infinite irreducible subalgebras of $\Cend_N$.

One of the results of the present paper is the classification of all subalgebras
 of \ce\,
and determination of the ones that act irreducibly on $\cp$ (Theorem 2.2).
This result proves the above-mentioned conjecture in the case $N=1$.  For
 general $N$ we can prove this conjecture only under the
assumption that the subalgebra in question is unital (see
Theorem~5.3).  This result is closely related to a difficult
theorem of A.~Retakh \cite{R} (but we avoid using it).

Next, we describe all finite irreducible  modules
over $\Cend_{N,P}$ (see Corollary~3.7).  This is
done by using the description of left ideals of the algebras
$\Cend_{N,P}$ (see Proposition~1.6a). Further, we describe all extensions
between non-trivial finite irreducible Cend$_{N,P}$-modules and between
 non-trivial finite irreducible and trivial finite dimensional modules
 (Theorem 3.10). This leads us to a
complete description of finite Cend$_{N}$-modules (Theorem 3.28).

Next we describe all automorphisms of Cend$_{N,P}$ (Theorems 4.2 and 4.3). We
 also classify all homomorphisms and anti-homomorphisms of Cend$_{N,P}$ to Cend$_N$ (Theorem 4.6). This gives, in particular, a classification of  anti-involutions of
Cend$_{N,P}$. One case of such an anti-involution ($N=1$,
$P=x$) was  studied by S.
Bloch [B] on the level of the Lie algebra of differential operators on the
 circle to link representations of the corresponding subalgebra to the  values
 of
$\zeta$-function. Representation theory of the subalgebra corresponding to the
 anti-involution of Cend$_{1}$ was developed in [KWY].

The subspace of anti-fixed points of an anti-involution of Cend$_{N,P}$ is a Lie
 conformal subalgebra that still acts irreducibly on $\Bbb C[\partial]^N$. This
 leads us to
Conjecture 6.20 on classification of infinite Lie conformal subalgebras of
gc$_N$  acting irreducibly on $\Bbb C[\partial]^N$. This conjecture agrees with
 the results of the papers [Z] and  [DeK].

We thank B. Bakalov for providing his results on the subject
of the paper, and we also thank D.~Djokovic for Theorem~4.25 and very useful
correspondance.
\


\head 1. Left and right ideals of Cend$_{N,P}$ \endhead

\vskip .3 cm

First we introduce the basic definitions and notations, see \cite{K1}.
An {\it   associative conformal algebra} $R$ is defined as a $\cp$-module
endowed with a $\Bbb C$-linear map,
$$
R\otimes R  \longrightarrow \Bbb C[\la]\otimes R, \qquad
a\otimes b  \mapsto a_\la b
$$
called the $\la$-product, and  satisfying the following axioms $(a,\, b,\, c\in
 R)$,

\

\noindent$(A1)_\la \qquad  (\p a)_\la b=-\la( a_\la b),\qquad  a_\la (\p
 b)=(\la+\p) (a_\la b)$

\

\noindent $(A2)_\la\qquad a_\la(b_\mu c)=(a_\la b)_{\la+\mu} c $

\

An associative conformal algebra is called $finite$ if it has finite rank as
 $\Bbb C[\p]$ -module. The notions of homomorphism, ideal and subalgebras of an
 associative conformal algebra are defined in the usual way.

A {\it  module} over an associative conformal algebra $R$ is  a $\cp$-module $M$
 endowed with a $\Bbb C$-linear map $
R\otimes M  \longrightarrow \Bbb C[\la]\otimes M$, denoted by  $
a\otimes v  \mapsto a_\la^M v$, satisfying the properties:
$$
\aligned
(\p a)_\la^M v &= [\p^M, a_{\la}^M]v= -\la( a_\la^M v),\quad  a\in R, \, v\in
 M,\\
a_\la^M(b_\mu^M v)&=(a_\la b)_{\la+\mu}^M v,\quad  a,b\in R.
\endaligned
$$
An $R$-module $M$ is called {\it trivial} if $a_{\lambda} v=0$
for all $a \in R$, $v \in M$ (but it may be non-trivial as a $\CC[\p]$-module).

Given two $\cp$-modules $U$ and $V$, a {\it conformal linear map} from $U$ to
 $V$ is a $\Bbb C$-linear map $a:U\to \Bbb C[\la]\otimes_{\Bbb C} V$, denoted by
 $a_{\la}:U\to V$, such that $[\p, a_{\la}]=-\la a_{\la}$, that is $\p^V a_\la -
 a_\la \, \p^U= -\la a_\la$. The vector space of all such maps, denoted by
 Chom$(U,V)$, is a $\cp$-module with
$$
(\p a)_{\la}:= -\la a_{\la}.
$$

 Now, we define Cend$V :=$ Chom$(V,V)$ and, provided that $V$ is a finite
 $\cp$-module, Cend$V$ has a canonical structure of an associative conformal
 algebra defined by
$$
(a_{\la}b)_{\mu} v = a_{\la}(b_{\mu-\la} v), \qquad \quad a,b\in \hbox{Cend }V,
 \ v\in V.
$$

\remark{Remark 1.1} Observe that, by definition, a structure of  a conformal
 module over an associative conformal algebra $R$ in a finite $\cp$-module $V$
 is the same as a homomorphism of $R$ to the associative conformal algebra
 Cend$V$. \endremark

For a positive integer $N$, let Cend$_N=$Cend$\, \cp^N$. It can also be viewed
as the associative conformal algebra associated to the associative algebra
Diff$^N \Bbb C^\times$ of all $N\times N$ matrix valued regular differential
operators on $\Bbb C^{\times}$, that is (see Sect. 2.10 in \cite{K1} for more
details)
$$
\hbox{Conf(Diff$^N \Bbb C^\times$)}=\oplus_{n\in \Bbb Z_+} \Bbb
 C[\partial]J^n\otimes Mat_N\Bbb
C
$$
with $\lambda$-product given by ($J^k_A=J^k\otimes A$)
$$
J^k_A\;_\lambda \; J^l_B=\sum_{j=0}^k \binom kj (\lambda+\partial)^j
J_{AB}^{k+l-j}.
$$

Given $\alpha \in \CC $, the natural representation of Diff$^N
\Bbb C^\times$ on $ e^{-\alpha t} \Bbb C^N[t, t^{-1}]$
gives rise a conformal module structure on $\cp^N$ over
Conf(Diff$^N \Bbb C^\times$), with $\la$-action

$$
J_A^m\,_\la\, v = (\la +\p + \alpha)^m Av,\quad m\in \Bbb Z_+, v\in \Bbb C^N.
$$

Now, using Remark 1.1, we obtain a natural homomorphism of conformal 
 associative algebras from Conf(Diff$^N \Bbb C^\times$) to Cend$_N$, which turns
 out to be an isomorphism (see [DK] and Proposition 2.10 in [K1]).

In order to simplify the notation, we will introduce the following bijective
 map, called the {\it symbol},
$$\aligned
\hbox{Symb}: \qquad \hbox{Cend}_N \qquad & \longrightarrow \hbox{Mat}_N \Bbb
 C[\partial, x]\\
\sum_k A_k(\partial)J^k &\longmapsto \sum_k A_k(\partial)x^k
\endaligned
$$
where $A_k(\partial)\in\hbox{Mat}_N(\Bbb C[\partial])$. The transfered
 $\lambda$-product is
$$
A(\partial,x)_\lambda B(\partial,x)=A(-\lambda,x+\lambda+\partial)
 B(\lambda+\partial,x)  .\tag 1.1
$$
The above $\la$-action of Cend$_N$ on $\Bbb C[\p]^N$ is given  by the following
 formula:
$$
A(\p,x)\,_\la\, v(\p)=A(-\la, \la +\p +\alpha ) v(\la+\p), \quad v(\p)\in \Bbb
 C[\p]^N.\tag 1.2
$$
Note also that under the change of basis of $\Bbb C[\p]^N$ by the invertible
 matrix $C(\p)$, the symbol $A(\p,x)$ changes by the formula:
$$
A(\p,x)\longmapsto C(\p+x) A(\p,x) C(x)^{-1}.\tag 1.3
$$

Observe that for any $C(x)\in $Mat$_N(\Bbb C[x])$, with non-zero constant
 determinant, the map (1.3) gives us an automorphism of Cend$_N$.

It follows immediately from the formula for $\lambda$-product that
$$
\hbox{Cend}_{P,N}:= P(x +\p) (\hbox{Cend}_N ) \quad \hbox{and}\qquad
 \hbox{Cend}_{N,P}:=  (\hbox{Cend}_N) P(x),
$$
with  $P(x)\in $ Mat$_N(\Bbb C[x])$, are right and left ideals, respectively, of
 Cend$_N$. Another important subalgebra  is
$$
Cur_N:= Cur\ (\hbox{Mat}_N) =\Bbb C[\p]\, (\hbox{Mat}_N \Bbb C).\tag 1.4
$$

\remark{Remark 1.5}  If $P(x)$ is nondegenerate, i.e.,~$\det P(x)
\neq 0$, then by elementary transformations over the rows (left
multiplications) we can make $P(x)$ upper triangular without
changing $\Cend_{N,P}$. After that, applying to Cend$_{N,P}$ an
automorphism of Cend$_N$ of the form (1.3), with det$\ C(x)=1$
(in order to multiply P on the right, which are elementary
transformations over the columns), we get
Cend$_{N,P}\simeq \hbox{Cend}_{N,D}$, with
$D=diag(p_1(x),\cdots,p_N(x))$, where $p_i (x)$ are monic
polynomials such that $p_i(x)$ divides $p_{i+1}(x)$. The $p_i
(x)$ are called the elementary divisors of $P$. So, up to conjugation, all
Cend$_{N,P}$ are parameterized by the sequence of elementary
divisors of $P$. \endremark

\

All left and right ideals of Cend$_N$ were obtained by B. Bakalov. Now, we
extend the classification to Cend$_{N,P}$.

\proclaim{Proposition 1.6} a) All left ideals in Cend$_{N,P}$,
with det$\ P(x)\neq 0$, are of the form Cend$_{N,QP}$, where
$Q(x)\in $ Mat$_N(\Bbb C[x])$.

b) All right ideals in Cend$_{N,P}$, with det$\ P(x)\neq 0$, are
of the form \break $Q(\p +x)$Cend$_{N,P} $, where $Q(x)\in
$Mat$_N(\Bbb C[x])$.

\endproclaim

\demo{Proof} (a) By Remark 1.5, we may suppose that $P$ is diagonal with det$\
 P(x)\neq 0$. Denote by $p_1(x),\dots , p_N(x)$ the diagonal coefficients.

Let $J\subseteq $Cend$_N$ be a left ideal. First, let us see that $J$ is
 generated over $\Bbb C[\p]$ by $I:=J\cap $Mat$_N(\Bbb C[x])$. If
 $a(\p,x)=\sum_{i=0}^m \p^i a_i(x)\in J$, then
$$
\aligned
E_{k,k}P(x)_\lambda a(\p,x) &= p_k(\lambda + \p +x) E_{k,k} a(\lambda + \p ,
 x)\\
&= p_k(\lambda + \p +x) E_{k,k}(\sum_i (\lambda + \p)^i a_i(x)) \in \Bbb
 C[\lambda]\otimes J,
\endaligned
\tag 1.7
$$
using that det$\ P(x)\neq 0$ and considering the maximal coefficient in
 $\lambda$ of (1.7),  we get $E_{k,k}a_m(x)\in J$ for all $k$. Hence $a_m(x)\in
 J$. Applying the same argument to $a(\p,x) - \p^m a_m(x)\in J$, and so on, we
 get $a_i(x)\in J$ for all $i$.  Therefore, $J$ is generated over $\Bbb C[\p]$
 by $I:=J\cap $Mat$_N(\Bbb C[x])$.

If $a(x)\in I$, then
$$
E_{i,j}P(x)_\lambda a(x)= p_j(\lambda +\p+x)E_{i,j} a(x)=\lambda^{max} E_{i,j}
 a(x)+ \hbox{lower terms}\in \Bbb C[\lambda]\otimes J.\tag 1.8
$$
Therefore,  Mat$_N (\Bbb C) \cdot I\subseteq I$.

Now, considering the next coefficient in $\lambda$ in (1.8)  if $p_j$ is
 non-constant, or the  constant term in $\lambda$ of  $xE_{i,j}P(x)_\lambda
 a(x)$  if $p_j$ is constant, we get that  $x a(x)\in I$. It follows that $I$ is
 a left ideal of Mat$_N(\Bbb C[x])$. But all left ideals of Mat$_N(\Bbb C[x])$
 are principal, i.e. of the form Mat$_N(\Bbb C[x]) R(x)$, since Mat$_N(\Bbb
 C[x])$ and $\Bbb C[x]$ are Morita equivalent. This completes the proof of (a).

In a similar way, but using the expression $a(\p, x)=\sum_i \p^i \tilde
 a_i(\p+x)$, we get (b).   \qed

\enddemo

\proclaim{Proposition 1.9}  Cend$_{N,P}\simeq\,B(\p+x)(\hbox{Cend}_N ) A(x)$ if
 $P(x)=A(x)B(x)$. In particular, Cend$_{N,P}\simeq\,$Cend$_{P,N}$.
 \endproclaim
 
 \demo{Proof} It is easy to see that the  map  $a(\p ,x)P(x)\to B(\p+x)a(\p
 ,x)A(x)$ is an isomorphism provided that $P(x)=A(x)B(x)$.\qed
 \enddemo

\vskip .3 cm


\head 2. Classification of  subalgebras of Cend$_1$ \endhead

\vskip .3 cm

We can identify \ce \ with $\cpx$, then the $\lambda-$product is
$$
r(\p,x)\ _{\lambda}\ s(\p,x)= r(-\la ,\la +\p +x) \ s(\la +\p , x), \tag 2.1
$$
where $r(\p,x),\ s(\p,x)\in\cpx $.

\vskip .3cm

The main result of this section is

\proclaim{Theorem 2.2} a) Any   subalgebra of \ce  \
is one of the following:
\roster
\item $\cp$;
\item $\cpx \ p(x)$, with $p(x)\in\Bbb C[x]$;
\item $\cpx \ q(\p +x)$, with $q(x)\in\Bbb C[x]$;
\item $\cpx \  p(x) \ q(\p +x)= \cpx \ p(x) \cap \cpx \ q(\p +x)$, with
$p(x),q(x)\in\Bbb C[x]$.
\endroster

\noindent b) The
subalgebras $\cpx \ p(x)$ with $p(x)\neq 0$, and  $\cp$  are all
the  subalgebras of Cend$_1$ that act irreducibly on $\cp$.
\endproclaim

\vskip .3cm

In order to prove Theorem 2.2, we first need some lemmas and  the following
 important notation. Given $r(\p , x)\in \cpx$, we denote by $r_i$ and $\tilde
 r_j$ the coefficients uniquely determined by
$$
r(\p,x)=\sum_{i=0}^n \ r_i(x) \p^i= \sum_{j=0}^m \ \tilde r_j(\p +x) \p^j\tag
 2.3
$$
with $r_n(x)\neq 0$ and $\tilde r_m(\p +x)\neq 0$.

\vskip .3cm

\proclaim{Lemma 2.4} Let $S$ be a subalgebra of \ce, let $p(x)$ and
$q(x)$ be two non-constant polynomials, and let $t(\p)\in \cp $ be a
non-zero polynomial.

\vskip .1cm

(a) If $t(\p)\in S $, then $\cp \subseteq S$.

\vskip .1cm

(b) If $t(\p), r(\p,x)\in S$ and $r(\p,x)$  depends non-trivially on $x$,
then $S=$\ce. In particular, if $1\in S$, then either $S=\cp$ or $S=$\ce.

\vskip .1cm

(c) If $p(x)\in S$, then $\cpx \ p(x)\subseteq S$.

\vskip .1cm

(d) If $q(\p + x)\in S$, then $\cpx \ q(\p + x)\subseteq S$.

\vskip .1cm

(e) If $p(x) q(\p+ x)\in S$, then $\cpx \ p(x) \ q(\p +x) \subseteq S$.

\endproclaim

\demo{Proof} (a) If $t(\p )\in S$, we deduce from the maximal coefficient in
 $\la$ of
$t(\p )\ _{\la}\  t(\p )= t(-\la )\  t(\la +\p)$
that $1\in S$, proving (a).

\vskip .2cm

(b) From (a), we have that $1\in S$. Then the coefficients of $\la$ in
$r(\p ,x)\ _{\la}\  1 = r(-\la ,\la +\p +x) $
are in $S$. Therefore, using notation (2.3), we obtain that
$\tilde r_j(\p +x)\in S$ for all $j$. Since $r(\p ,x)$ depends non-trivially
on $x$, there exist $j_0$ such that $\tilde r_{j_0}$ is non-constant, that is
$\tilde r_{j_0}(z)=\sum_{i=0}^l a_i z^i$ with $a_l\neq 0$ and $l>0$. Now,
using that $\cp \subseteq S$ and
$$
1\ _{\la}\  \tilde r_{j_0} (\p +x)= \tilde r_{j_0} (\la +\p +x)\, = \, \la^l +
 \left(l \, a_l (\p +x) + a_{l-1}\right)\  \la^{l-1} + \hbox{ lower powers in
 }\la
$$
we obtain that $x\in S$. Then by induction and  taking $\la$-products of type
$x_{\la} x^k$ we see that $x^{k+1}\in S$ for all $k\geq 1$, proving (b).

\vskip .2cm

(c) Let $p(x)=\sum_{i=0}^n a_i x^i$, with $a_n\neq 0$ and $n>0$. Then,
 considering the coefficient of $\la^{n-1}$ in $p(x)\ _{\la} \ p(x)= p(\la +\p
 +x) p(x)$, we get that\break
$(n \, a_n \, (\p +x) \, + \, a_{n-1}) \, p(x)\in S$. Since $S$ is a
 $\cp$-module, we have $\p p(x)\in S$, obtaining that $x\, p(x)\in S$. Applying 
 this argument to $  x\, p(x)$, we get that $x^2 p(x)\in S$, etc, and   $x^k \,
 p(x)\in S$ for all $k>0$, proving (c).

\vskip .2cm

(d) The proof is identical to that of (c).

\vskip .2cm

(e) Assume that $q(x+\p)p(x)\in S$. Then, we compute $q(x+\p)p(x)_\la
q(x+\p)p(x)=q(x+\p)p(\la+\p+x)q(\la+x+\p)p(x)$, and looking at the
monomial of highest degree minus one, we get that $(x+\p) q(x+\p)
p(x)\in S$, and since by definition $S$ is a $\cp$-module, we deduce
that $q(x+\p) \tilde p(x):=x q(x+\p)p(x)\in S$. Applying this argument
to $q(x+\p) \tilde p(x)$ we deduce that $x^k q(x+\p)p(x)\in S$ for any
$k\in \Bbb Z_+$, and therefore $q(x+\p)p(x)\cpx\subseteq S$.\qed
\enddemo

\vskip .3cm

\vskip .3cm

\proclaim {Lemma 2.5} Let $S$ be a subalgebra of \ce which does not contain $1$.

\vskip .1cm

(a) Let $p(x)$ be of minimal degree such that $p(x)\in S$. Then $\cpx p(x)= S$.

\vskip .1cm

(b) Let $q(\p+x)$ be of minimal degree such that $q(\p+x)\in S$. Then $S=\cpx
 q(\p+x)$.

\vskip .1cm

(c)  Let $q(\p+x)p(x)$ be of minimal degree (in x) such that $q(\p+x)p(x)\in S$.
 Then $S=p(x)q(\p+x)\cpx$.

\endproclaim

\vskip .3cm

\demo{Proof} (a) From Lemma 2.4.(c), we have that $p(x)\cpx \subseteq S$
(by our assumption, $p(x)$ is non-constant). Now,  suppose that there exist
$q(\p,x)\in S$ with $q(\p,x)\notin p(x)\cpx$ and $p$ as above. Then, by
applying the division algorithm to each coefficient of $q(\p,x)=\sum_{k=0}^l
 q_k(x)\p^k$, we may write $q(\p,x)=t(\p,x)p(x)+r(\p,x)$ with  $r(\p,x)=
 \sum_{k=0}^n r_k(x)\p^k=\sum_{j=0}^m\tilde r_j(\p+x)\p^k$ and  $\deg r_k<\deg
 p$ (cf. notation (2.3)). Using that $p(x)\cpx \subseteq S$, we obtain  that
 $r(\p,x)\in S$. Now, since
$$
r(\p,x)\ _\la\ r(\p,x)=r(-\la,\la+\p+x)r(\la+\p,x),\tag 2.6
$$
looking at the coefficient of maximum degree in $\la$ in (2.6), we get:
$r_n(x)\tilde r_m(x+\p)\in S$. By our assumption, one of the polynomials
in this product is non-constant.
If $\tilde r_m(x+\p)$ is constant, then $r_n(x)\in S$, but $\deg r_n<\deg p$
which is a contradiction. If $r_n(x)$ is constant, then $\tilde r_m(x+\p)\in S$.
 Then, looking at the leading coefficient of the following polynomial in
$\la$: $p(x)_\la\ \tilde r_m(x+\p)=p(\la+\p+x)\tilde r_m(x+\la+\p)$ we have that
 $1\in S$, which contradicts our assumption.

If neither $\tilde r_m(x+\p)$ nor $r_n(x)$ are constants, we look at
$p(x)_\la\  \tilde r_m(x+\p) r_n(x)=p(\la+\p+x)\tilde r_m(\la+x+\p) r_n(x)\in
S$ and looking at the coefficient of maximum degree in $\la$ we get that
 $r_n(x)\in S$, which contradicts the minimality of $p(x)$.

\

(b) The proof is the same as that of (a).

\

(c) We may assume that $p$ and $q$ are non-constant polynomials, otherwise we
are in the cases (a) or (b). By Lemma 2.4(e), we have $p(x)q(x+\p)\cpx \subseteq
 S$.   Let $t(\p,x)\in S$, but $t(\p,x)\notin \cpx p(x)q(x+\p)$. Then we
may have three cases:

(1) $t(\p,x)\in p(x)\cpx$ or

(2) $t(\p,x)\in q(\p+x)\cpx$ or

(3) $t(\p,x)\notin p(x)\cpx$ nor $t(\p,x)\notin q(\p+x)\cpx$.

Note that these cases are mutually exclusive. Suppose we are in Case (1),
so that $t(\p,x)=p(x)r(\p,x) $ with $r(\p,x)\notin q(\p+x)\cpx$. Then we get $r(\p,x)=
 q(\p+x)\tilde r(\p,x)+s(\p,x)$, with $s(\p,x)\neq 0$, and (using notation
 (2.3)) $\deg\ \tilde s_k<\deg q$ for all $k=0,\dots,m$. Therefore, we have that
$t(\p,x)=p(x)r(\p,x)=\break p(x)q(\p+x)\tilde r(\p,x)+p(x)s(\p,x)$ and then
$p(x)s(\p,x)\in S$. Now, we can compute:
$$
p(x)s(\p,x)_\la\ p(x)q(\p+x)=p(\la+\p+x)s(-\la,\la+\p+x)p(x)q(\la+\p+x)
$$
and looking at the coefficient of maximum degree in $\la$, we have (using
 notation (2.3)) that $p(x)\tilde s_m(\p+x)\in S$ which is a contradiction.

\smallpagebreak

Similarly, Case (2) also leads to a contradiction.

\smallpagebreak

In the remaining Case (3) we may assume that $\deg p\leq \deg q$ since the
case of the opposite inequality is completely analogous.
We have $t(\p,x)\in S$, but  $\notin \cpx p(x)$. Then
$$t(\p,x)=p(x)h(\p,x)+r(\p,x)\tag 2.7$$
with $0\neq r(\p,x)= \sum_{k=0}^n r_k(x)\p^k=\sum_{j=0}^m\tilde r_j(\p+x)\p^k$
 where $\deg   r_k<\deg p$ and $\deg   \tilde r_j<\deg p$.

If $h(\p,x)\in\cpx q(\p+x)$, then $r(\p,x)\in S$, but the leading coefficient of
$$
p(x)q(\p+x)\ _\la\ r(\p,x)=p(\la+\p+x)q(\p+x) r(\la+\p,x)
$$
is in $S$ which is $q(\p+x)r_n(x)$, and this contradicts the assumption of
 minimality of $p(x)q(\p+x)$.

So, suppose that $h(\p,x)\notin\cpx q(\p+x)$. Then $h(\p,x)= \tilde
 h(\p,x)q(\p+x)+s(\p,x)$ with $0\neq s(\p,x)= \sum_{k=0}^l
 s_k(x)\p^k=\sum_{j=0}^m\tilde s_j(\p+x)\p^k$ and $\deg   \tilde s_j<\deg q$. By
 (2.7) we have $p(x)s(\p,x)+r(\p,x)\in S$. Now, we compute:
$$
\aligned
\left(p(x)s(\p,x)\right.+& \left. r(\p,x)\right)\ _\la \ p(x)q(\p+x)\\
&\!\!\!\!\!\!\!\!\!\!=\left(p(\la+\p+x)s(-\la,\la+\p+x)+r(-\la,\la+\p+x)\right)
 p(x)q(\la+\p+x)
\endaligned
$$
Then the leading coefficient  in $\la$ is either $p(x)\tilde s_m(\p+x)\in S$,
which  is impossible since $\deg \tilde  s_m<\deg q$, or
$p(x)\tilde r_m(\p+x)\in S$. But in the latter case, $\deg \tilde r_m\geq \deg
 q$, but by construction $\deg \tilde r_m< \deg p$, and this contradicts the
 assumption $\deg p\leq \deg q$.\qed

\enddemo

\demo{Proof of Theorem 2.2} (a) Let $S$ be a non-zero subalgebra of \ce.
If $S\subseteq \cp$ then by Lemma 2.4.(a) we have that $S=\cp$. Therefore we
may assume that there is $r(\p,x)\in S$ which depends nontrivially on x.
Recall that we can write
$
r(\p,x)=\sum_{i=0}^m p_i(x)\p^i=\sum_{j=0}^n q_j(\p+x)\p^j.$
We have
$$\aligned
r(\p,x)\ _\la\  r(\p,x)& =r(-\la,\la+\p+x)r(\la+\p,x)=\\
& =\sum_{i=0}^m  \sum_{j=0}^n q_j(\p+x)p_i(x)(-\la)^j (\la+\p)^i
\endaligned
$$
Then, considering the leading coefficient of this $\la$-polynomial,   we
have\break
 $p_m(x) q_n(\p+x)\in S$. Therefore, we may have one of the following
 situations:

\roster
\item  $p_m(x)$ and $q_n(\p+x)$ are constant,

\item  $q_n(\p+x)$ is  constant and $p_m(x)$ is non-constant,

\item  $p_m(x)$ is constant and $q_n(\p+x)$ is non-constant, or

\item  both polynomials non-constant.
\endroster

\smallpagebreak

\noindent Let us see what happens in each case:

\smallpagebreak

\noindent (1) By Lemma 2.4.(b), we have that $S=$\ce.

\smallpagebreak

\noindent (2) In this case, we  may take $p(x)\in S$ of  minimal degree, then
 using Lemma 2.5.(a) we have  $S=\cpx p(x) $.

\smallpagebreak

\noindent (3) It is completely analogous to (2).

\smallpagebreak

\noindent (4) Here, we have that $p(x)q(x+\p) \in S$ and, again we may assume
that it has minimal degree. Now, by  Lemma 2.5.(c), we finish the proof of
(a).

The proof of (b) is straightforward.\qed\enddemo

\vskip .3 cm

%
%

\head 3. Finite  modules over  Cend$_{N,P}$ \endhead

\vskip .3 cm

Given $R$ an associative conformal algebra (not necessarily
finite), we will establish a correspondence between the set of
maximal left ideals of $R$ and the set of irreducible
$R$-modules. Then we will apply it to the subalgebras
Cend$_{N,P}$.

First recall that the following property holds in an  $R$-module $M$ (cf.~Remark
 3.3 [DK]):
$$ a_\lambda (b_{-\p-\mu}v)=(a_\lambda b )_{-\p-\mu} v \qquad
\quad a,\, b\in R, \ v\in M .\tag 3.1
$$

\remark{Remark 3.2} (a) Let $v\in M$ and fix $\mu\in \Bbb C $,
then due to (3.1) we have that $R_{-\p-\mu}v$ is an $R$-submodule
of $M$.

(b) $Tor\, M$  is a trivial $R$-submodule of $M$ (Lemma 8.2, [DK]).

(c) If $M$ is irreducible and $M=Tor\, M$, then  $M\simeq \Bbb C$.

(d) If $M$ is a non-trivial finite irreducible $R$-module, then $M$ is free as a
 $\CC [\partial]$-module.
\endremark

\vskip .2cm

\proclaim{Lemma 3.3} Let $M$ be a non-trivial irreducible
$R$-module.  Then there exists $v\in M$ and $\mu\in\Bbb C$ such
that $R_{-\p -\mu} v\neq 0$.  In particular,  $R_{-\p -\mu} v=M$
if $M$ is irreducible.
\endproclaim

\demo{Proof} Suppose that $R_{-\p -\mu} v=0$ for all $v\in M$ and
$\mu\in \Bbb C$, then we have that $r_{-\p -\mu} v=0$ in $\Bbb
C[\mu]\otimes M$ for all $r\in R$ and $v\in M$. Thus writing down
$r_{-\p -\mu} v$ as a polynomial in $\mu$ and looking at the
$n$-products that are going to appear in this expansion, we
conclude that $r_\lambda v=0$ for all $v\in M$ and $r\in
R$. Hence $M$ is a trivial $R$-module,  a contradiction.
 \qed
\enddemo

\

By Lemma 3.3, given a non-trivial irreducible $R$-module $M$  we can fix $v\in M
 $ and $\mu\in \Bbb C$ such that $R_{-\p -\mu} v=M$ and consider the following
 map
$$
\phi: R\to M , \qquad r \mapsto r_{-\p -\mu} v.
$$
Observe that $\phi (\p r)=(\p + \mu) \, \phi(r)$ and using (3.1) we also have
 $\phi(r_\lambda s)=r_\lambda \phi(s)$. Therefore,  the map $\phi$ is a
 homomorphism of $R$-modules into $M_{-\mu}$,  where $M_{\mu}$ is the
 $\mu$-twisted module of $M$ obtained by replacing $\p$ by $\p +\mu$ in the
 formulas for the action of $R$ on $M$, and Ker$(\phi)$ is a maximal  left ideal
 of $R$. Clearly this map is onto $M_{-\mu}$.

Therefore we have that $M_{-\mu} \simeq (R/\hbox{Ker }\phi)$ as $R$-modules, or
 equivalently,
$$
M\simeq (R/\hbox{Ker }\phi)_{\mu}.\tag 3.4
$$
On the other hand, it is immediate that given any maximal left ideal I of $R$, 
 we have that $(R/I)_\mu$ is an irreducible $R$-module. Therefore we have proved
 the following

\proclaim{Theorem 3.5} Formula (3.4) defines a surjective map from the set of
 maximal left ideals of $R$ to the set of equivalence classes of non-trivial
 irreducible $R$-modules.
\endproclaim

\remark{Remark 3.6} (a) Observe that given an $R$-module $M$ and $v\in M$, the
 set $I=\{a\in R \, |\,  a_\lambda v=0\}$ is a left ideal, but not necessarily
 $M\simeq R/I$. For example, consider $\Bbb C[\p]$ as a Cend$_1$-module, then
 the kernel of $a\mapsto a_\lambda v$ is $\{0\}$.

(b) If we fix $\mu\in \Bbb C$, there are examples of irreducible modules where
 $R_{-\p -\mu} v=0$ for all $v\in M$ (cf. Lemma 3.3). Indeed, consider $\Bbb
 C[\p ]$ as a Cend$_{1, (x+\mu)}$-module.
\endremark

\

Using Remark 3.2, Proposition 1.6  and Theorem 3.5, we have

\proclaim{Corollary 3.7} The Cend$_{N,P}$-module $\Bbb C[\p]^N$
defined by (1.2) is irreducible if and only if  det$\, P(x) \neq
0$. These are all non-trivial irreducible Cend$_{N,P}$-modules up
to equivalence, provided that $\det P(x) \neq 0$.
\endproclaim

Note that Corollary~3.7 in the case $P(x) =I$, have been
 established earlier in \cite{K2}, by a completely different
method (developed in \cite{KR}).

\

A subalgebra $S$ of Cend$_N$  is called $irreducible$ if $S$
acts irreducibly in $\Bbb C[\p]^N$.

\proclaim{Corollary 3.8} The following   subalgebras of
Cend$_N$ are irreducible: Cend$_{N,P}$ with
det$\,P(x)\neq\,0$, and Cur$_N$ or conjugates of it by automorphisms (1.3).
\endproclaim

\

\remark{Remark 3.9} It is easy to show that every non-trivial
irreducible representation of $Cur_N$ is equivalent to the
standard module $\CC [\partial]^N$, and that every finite
module over $Cur_N$ is completely reducible.
\endremark

\

We will finish this section with the classification of all extensions of
 Cend$_{N,P}$-modules involving the
standard module $\Bbb C[\p]^N$ and
finite dimensional trivial modules, and the classification of all finite modules
over Cend$_N$.

We shall work with the standard irreducible
Cend$_{N,P}$-module  $\Bbb C[\p]^N$ with $\lambda$-action (see (1.2))
$$
a(\p,x)P(x)_\lambda v(\p)= a(-\la , \la +\p + \alpha) P(\la +\p) v(\la +\p).
$$

Consider  the trivial
Cend$_{N,P}$-module over the finite dimensional vector space $V_T$, whose $\Bbb
C [\p ]$-module structure is given by the linear operator $T$, that is: $\p
\cdot v = T(v)$, $v\in V_T$. As usual, we may assume that
$P(x)=diag\{p_1(x),\cdots , p_N(x)\}$. We shall assume that $\det P \neq
0$.

\proclaim{Theorem 3.10} a) There are no non-trivial extensions of
Cend$_{N,P}$-modules of the form:
$$
0 \to V_T \to E \to \Bbb C [\p ]^N \to 0.
$$

\noindent b) If there exists a non-trivial extension of
Cend$_{N,P}$-modules of the form
$$
0\to \CC[\p]^N  \to E \to V_T \to 0, \tag 3.11
$$
then $\det P(\alpha +c)=0$ for some eigenvalue $c$ of $T$. 
In this case, all  torsionless 
extensions of $\CC[\p]^N$ by finite dimensional vector spaces, are parameterized
by decompositions $P(x +\alpha )=R(x)S(x)$ and can be realized as follows. 
 Consider  the following isomorphism of conformal algebras:
$$
\Cend_{N,P} \to S(\p+x)\Cend_N R(x),\quad a(\p,x)P(x)\mapsto S(\p+x)a(\p,x)
R(x),
$$
where $P(x +\alpha)=R(x)S(x)$, (this is the
isomorphism between Cend$_{N,S}$ and Cend$_{S,N}$ (Proposition 1.9), restricted to
$\Cend_{N,R}S(x)$). 
Using this isomorphism, we get an action of Cend$_{N,P}$ on
$\CC[\p]^N$:
$$
a(\p,x)P(x)_\la v(\p)=S(\p)a(-\la,\la+\p +\alpha)R(\la+\p) v(\la+\p).
$$
Then $S(\p)\CC[\p]^N$ is a submodule isomorphic to the standard module,
of finite codimension in $\CC[\p]^N$. 

\noindent c) If $E$ is a  non-trivial extension of
Cend$_{N,P}$-modules of the form:
$$
0 \to \Bbb C [\p ]^N  \to E \to \Bbb C [\p ]^N \to 0,
$$
then $E=\Bbb C [\p ]^N \otimes \Bbb C^2$ as a $\CC[\p]$-module (with trivial action of $\p$ on $\CC^2$) and Cend$_{N,P}$ acts by
$$
a(\p , x)_{\lambda} (c(\p) \otimes u)=  a(-\lambda , \lambda +  \p \otimes 1
 + 1\otimes J ) c(\lambda + \p)( 1\otimes u),\tag 3.12
$$
where $J$ is a $2\times 2$ Jordan block matrix.
\endproclaim

\demo{Proof} a) Consider a short exact sequence of
$R=$Cend$_{N,P}$-modules
$$
0\to T\to E\to V\to 0,\tag 3.13
$$
where $V$ is irreducible finite, and $T$ is trivial (finite dimensional vector
space). Take $v\in  E$ with  $v\notin T$, and  let $\mu\in \Bbb C$  be such that
$A:=R_{-\p -\mu}v\neq 0$. Then we have three  possibilities:

1) The image of $A$ in $V$ is $0$, then $A=T$, which is impossible since $A$
corresponds to a left ideal of Cend$_{N,P}$.

2) The image of $A$ in $V$ is $V$ and $A\cap T=0$, then $A$ is isomorphic to
$V$, hence the exact sequence splits.

3) The image of $A$ in $V$ is $V$ and $T'=A\cap T\neq 0$. Now, if $T'=T$ then
$A=E$ and $E$ is a cyclic module, which is impossible since it has torsion. If
$T'\neq T$, we consider the exact sequence $0\to T'\to A\to V\to 0$, by an
inductive argument on the dimension of the trivial module, the last sequence
split, i.e. $A=T'\oplus V'\subset E$ with $V'\simeq V$, hence $E=T\oplus V'$ as
Cend$_{N,P}$-modules, proving (a).

\

b) We may assume without loss of generality that $\alpha=0$. Consider an extension of
Cend$_{N,P}$-modules of the form (3.11).
As a vector space $E=\CC[\p]^N\oplus V_T$. We have, for $v\in V_T$:
$$
\aligned
\p\ v &= T(v) + g_v(\p), \, \hbox{ where } g_v(\p)\in \CC[\p]^N, \\
x^l BP(x)_\la v &= f_l^{v,B} (\la ,\p),\, \hbox{ where } f_l^{v,B} (\la ,\p)\in
(\CC [\p]^N)[\la], \ B\in \Mat_N\CC.
\endaligned \tag 3.14
$$
Let $P(x)=\sum_{i=0}^m Q_i x^i$. Since
$$
\aligned
(x^k AP(x)_\la x^lBP(x))_{\la+\mu} v &= (\la+\p+x)^k AP(\la+\p+x)x^lBP(x)_{\la+\mu}
v \\
&= \sum_{i=0}^m\sum_{j=0}^{i+k} \binom{i+k}{j}(\la+\p)^{i+k-j} x^{j+l}
AQ_iBP(x)_{\la+\mu} v\\
&= \sum_{i=0}^m\sum_{j=0}^{i+k}
\binom{i+k}{j}(-\mu)^{i+k-j} f_{j+l}^{v,AQ_iB}(\la+\mu,\p)
\endaligned
$$
and
$$
\aligned
x^k AP(x)_\la (x^lBP(x)_{\mu} v) &= x^k AP(x)_\la (f_l^{v,B} (\mu ,\p))\\
&= (\la+\p)^{k}AP(\la+\p) f_{l}^{v,B}(\mu,\la+\p)
\endaligned
$$
must be equal by (A2)$_\la$, we have the functional equation
$$
(\la+\p)^{k}AP(\la+\p) f_{l}^{v,B}(\mu,\la+\p)=\sum_{i=0}^m\sum_{j=0}^{i+k}
\binom{i+k}{j}(-\mu)^{i+k-j} f_{j+l}^{v,AQ_iB}(\la+\mu,\p).\tag 3.15
$$
If we  put $\mu=0$ in (3.15), we get
$$
(\la+\p)^{k}AP(\la+\p) f_{l}^{v,B}(0,\la+\p)=\sum_{i=0}^m
f_{i+k+l}^{v,AQ_iB}(\la,\p).\tag 3.16
$$
Since the right-hand side of (3.16) is symmetric in $k$ and $l$, so is the left-hand side, hence, in particular, we
have
$$
(\la+\p)^{k}AP(\la+\p) f_{0}^{v,B}(0,\la+\p)= AP(\la+\p)
 f_{k}^{v,B}(0,\la+\p).
$$
Taking $A=I$ and using that $\det P\neq 0$, we get
$$
f_{k}^{v,B}(0,\la+\p)=(\la+\p)^{k} f_{0}^{v,B}(0,\la+\p).\tag 3.17
$$
Furthermore, by (A1)$_\la$, we have $[\p, x^k AP(x)_\la ] v=-\la \ x^k AP(x)_\la v$, which gives us the next condition:
$$
(\la+\p)f_{k}^{v,A}(\la,\p)=f_{k}^{T(v),A}(\la,\p)+(\la+\p)^{k}AP(\la+\p)
g_v(\la+\p).\tag 3.18
$$
 
We shall prove that if $c$ is an eigenvalue of $T$ and $p_j(c)\neq 0$  for all $1\leq j\leq N$, then (after a change of complement) the generalized eigenspace of $T$ corresponding to the eigenvalue $c$  is a trivial submodule of $E$ (hence is a non-zero torsion submodule). Indeed, 
let $\{v_1,\cdots , v_s\}$ be  vectors corresponding to one Jordan block of $T$ associated to $c$, that is $T(v_1)=cv_1$ and $T(v_{i+1})=c v_{i+1} +v_i$ for $i\geq 1$. Then (3.18) with  $v=v_1$ becomes
$$
(\la+\p-c)f_{k}^{v_1,A}(\la,\p)=(\la+\p)^k AP(\la+\p)
g_{v_1}(\la+\p)\tag 3.19
$$
Observe that the right-hand side of (3.19) depends on $\la+\p$, so $f_{k}^{v_1,A}(\la,\p)=f_{k}^{v_1,A}(0,\la+\p)$.  Then using (3.17), we have
$$
f_{k}^{v_1,A}(\la,\p)=f_{k}^{v_1,A}(0,\la+\p)=(\la+\p)^{k} f_{0}^{v_1,A}(0,\la+\p)=
(\la+\p)^{k} f_{0}^{v_1,A}(\la,\p)\tag 3.20
$$
Similarly, considering (3.18) with  $v=v_{i+1}$ ($i\geq 1$), we get
$$
\aligned
(\la+\p-c)f_{k}^{v_{i+1},A}(\la,\p)&=f_{k}^{v_{i},A}(\la,\p)+(\la+\p)^k AP(\la+\p)
g_{v_{i+1}}(\la+\p)\\
&=(\la+\p)^k \left[f_{0}^{v_i,A}(0,\la+\p) + AP(\la+\p)
g_{v_{i+1}}(\la+\p)\right]
\endaligned
\tag 3.21
$$
Again, since the right hand side of (3.21) depends only on $\la +\p$, we have that (3.20) also holds for any $v_i$.

Using that $p_j(c)\neq 0$ ($j=1,\cdots, N$) (recall that $P$ is diagonal), 
and taking $A=E_{i,l}$, we obtain from (3.19) with $k=0$ that
$$
f_{0}^{v_1,A}(\la,\p)=AP(\la+\p)
h_{v_1}(\la+\p)\tag 3.22.a
$$
where $g_{v_1}(\p)=(\p-c)h_{v_1}(\p)$. Now, (3.21) with $k=0$ and $i=1$ becomes (by (3.22.a)) 
$$
\aligned
(\la+\p-c)f_{0}^{v_2,A}(\la,\p)&=f_{0}^{v_1,A}(\la,\p)+AP(\la+\p)
g_{v_2}(\la+\p)  \\
&=  AP(\la+\p)\left(h_{v_1}(\la+\p) +
g_{v_2}(\la+\p)\right)
\endaligned
$$
As in (3.22.a), we get
$$
f_{0}^{v_2,A}(\la,\p)=AP(\la+\p)
h_{v_2}(\la+\p)
$$
where $g_{v_2}(\p)+ h_{v_1}(\p)=(\p-c)h_{v_2}(\p)$. Similarly, we obtain for all $i\geq 1$,
$$
f_{0}^{v_{i+1},A}(\la,\p)=AP(\la+\p)
h_{v_{i+1}}(\la+\p)\tag 3.22.b
$$
where $g_{v_{i+1}}(\p)+ h_{v_i}(\p)=(\p-c)h_{v_{i+1}}(\p)$. Changing the basis to $v'_i=v_i-h_{v_i}(\p)$, we have from (3.22.a-b) that $x^kAP(x)_\la v'_i=0$ and 
$$
\aligned \p v'_1
&=T(v_1)+ g_{v_1}(\p)-\p h_{v_1}(\p)\\
&= cv_1 +(\p-c)h_{v_1}(\p)-\p h_{v_1}(\p)=cv'_1,
\endaligned
$$
$$
\aligned
\p v'_{i+1}
&=T(v_{i+1})+ g_{v_{i+1}}(\p)-\p h_{v_{i+1}}(\p)\\
&= cv_{i+1} +v_i  +(\p-c)h_{v_{i+1}}(\p)-\p h_{v_{i+1}}(\p)- h_{v_{i}}(\p)\\
&=cv'_{i+1}+v'_i.
\endaligned\tag 3.23
$$
Hence, the $T$-invariant subspace spanned by $\{v'_i\}$ is a trivial submodule of $E$. 
Therefore, if  $p_j(c)\neq 0$ for all $j$ and all eigenvalues $c$  of $T$, then $E$ is a trivial extension. This proves the first part of (b).

Now suppose that the extension $E$ of $\CC[\p]^N$ by a finite dimensional vector space   have no non-zero trivial submodule (equivalently, $E$ is torsionless). By Remark 3.2.(b), $E$ must be a free $\CC[\p]$-module of rank $N$.

Then, the problem reduces to the study of a Cend$_{N,P}$-module structure on $E=\CC [\p]^N$, but using Remark 1.1, this is the same as a non-zero homomorphism  from Cend$_{N,P} $ to Cend$_N$. So, the end of this proof also gives us the classification of all these homomorphisms. 

Denote by $\phi: \ $Cend$_{N,P}\to
$Cend$_N$ the (non-zero) homomorphism associated to $E$. It is an embedding (due to irreducibility) of free $\CC[\p]$-modules 
 $\Bbb C[\p]^N\to 
 \Bbb C[\p]^N$, hence it is given by a non-degenerate matrix $S(\p)\in\  $Mat$_N \CC [\p]$. Hence the action on $E$ of Cend$_{N,P}$ is given by the formula:
$$
\phi(a(\p,x)P(x))\,_\lambda\, (S(\p)v)=S(\p)\, a(-\lambda,\lambda+\p+\alpha)\,P(\lambda+\p+\alpha)\,v\qquad\hbox{ for all }
v\in \Bbb C^N.
$$
Furthermore, we have:
$$
\aligned
\left(\phi(a(\p,x)P(x))\,S(x)\right)\,_\lambda \,v &= \phi(a(\p,x)P(x))\,_\lambda \,(S(\p)v )\\
& =(S(\p+x)a(\p, x+\alpha)P(x+\alpha))\,_\lambda \,v \qquad\hbox{ for all }
v\in \Bbb C^N.
\endaligned
$$
Hence $\phi(a(\p,x)P(x))=S(\p+x)\,a(\p, x+\alpha)\,P(x+\alpha)\,S^{-1}(x)$, and this is in Cend$_N$ if and only if
$R(x):=P(x+\alpha)\,S^{-1}(x)\in \hbox{Mat}_N\Bbb C[x]$, proving b).

\

c) Consider a short exact sequence of $R=$Cend$_{N,P}$-modules
$$
0\to V\to E\to V'\to 0,\tag 3.26
$$
where $V$ and $V'$ are irreducible finite. Take $v\in  E$ with  $v\notin V$, and
 let $\mu\in \Bbb C$  be such that $A:=R_{-\p -\mu}v\neq 0$. Then we have three
 possibilities:

1) The image of $A$ in $V'$ is $0$, then $A=V$, which is impossible because
$v\notin V$.

2) The image of $A$ in $V'$ is $V'$ and $A\cap V=0$, then $A$ is isomorphic to
$V'$, hence the exact sequence splits.

3) The image of $A$ in $V'$ is $V'$ and $A\cap V=V$, hence $A=E$ and $E$ is a
cyclic module, hence corresponds to a left ideal which is contained in a unique
max ideal (otherwise the sequence splits).
It is easy to see then that $E$ is the indecomposable
module given in (3.12), where  $J$ is the $2\times 2$ Jordan block.\qed

\enddemo

\proclaim{Corollary 3.27} There are no non-trivial extensions of
Cend$_{N}$-modules of the form:
$$
0 \to V_T \to E \to \Bbb C [\p ]^N \to 0 \quad \hbox{ or }\quad 0 \to
\Bbb C [\p ]^N \to E \to  V_T\to 0
$$
\endproclaim

\proclaim{Theorem 3.28} Every finite
 Cend$_{N}$-module is isomorphic to a direct sum of its (finite dimensional) trivial torsion submodule  and a free finite $\CC[\p]$-module $\Bbb C[\p]^N\otimes T$ on which the
 $\lambda$-action is given by
$$
a(\p , x)_{\lambda} (c(\p) \otimes u)=  a(-\lambda , \lambda +  \p \otimes 1
 + 1\otimes \alpha ) c(\lambda + \p)( 1\otimes u), \tag 3.29
$$
where $\alpha$ is an arbitrary operator on  $T$.
\endproclaim

\demo{Proof} Consider a short exact sequence of $R=$Cend$_{N}$-modules
$$
0\to V\to E\to V'\to 0,
$$
where $V$ and $V'$ are irreducible finite. By Theorem 3.10(c), the exact
sequence split or  $E$ is the indecomposable module that corresponds to a
$2\times 2$ Jordan block $J$, i.e.,~$E=\Bbb C [\p ]^N\otimes \Bbb C^2$, and
$R$ acts via (3.29), where $S=0$, $\beta = 0$ and $\alpha
=J$.

Next, using Corollary 3.27, the short exact
sequences of $R$-modules $0\to V \to E\to C\to 0$  and $0\to C
\to E \to V \to 0$, where $C$ is a trivial $1$-dimensional
$R-$module, and $V$ is a standard $R$-module (1.2),
split.

Recall \cite{K1} that an $R$-module is the same as a module over
the associated extended annihilation algebra $(\Alg R)^- = \CC
\partial \ltimes (\Alg R)_-$, where $(\Alg R)_-$ is the
annihilation algebra.  For $R = \Cend_{N}$ one has:
$$
(\Alg R)_- = (\Diff^N \CC)\, ,
(\Alg R)^- =\CC \partial \ltimes (\Alg R)_- \, ,
$$
where $\partial$ acts on $(\Alg R)_-$ via $-ad \partial_t$.
Furthermore, viewed as an $(\Alg R)_-$-module, all modules (1.2)
are equivalent to the  module $F= \CC [t,t^{-1}]^N / \CC [t]^N$ , and
the modules (1.2) are obtained by letting $\partial$ act as
$-\partial_t + \alpha$.

Let $M$ be a finite $R$-module.  Then it has finite length and,
by Corollary~3.7, all its irreducible subquotients are either
trivial $1$-dimensional or are isomorphic to a standard
$R$-module (1.2).  Since in the case~b) above, the exact sequence
splits when restricted to $(\Alg R)_-$, we conclude that, viewed
as an $(\Alg R)_-$-module, $M$ is a finite direct sum of modules
equivalent to $F$ or trivial $1$-dimensional.  Thus, viewed as an
$(\Alg R)_-$-module, $M =S \oplus (F \otimes T)$, where $S$ and $T$
are trivial $(\Alg R)_-$-modules.  The only way to extend this
$M$ to an $(\Alg R)^-$-module is to let $\partial$ act as
operators $\alpha$ and $\beta$ on $T$
 and $S$, respectively, and as $-\partial_t$ on $F$,
which gives (3.29).   \qed

\enddemo

\vskip .3 cm

%
%

\head 4. Automorphisms and anti-automorphisms of Cend$_{N,P}$ \endhead

\vskip .3 cm

A $\cp$-linear map $\s:R\to S$ between two associative conformal
algebras is called a {\it homomorphism  }({\it resp. anti-homomorphism}) if
$$
 \sigma (a_{\lambda}b)= \sigma (a)_{\lambda} \sigma (b) \quad
        (\hbox{resp  }\s(a_\la b)= \s(b)_{-\la-\p} \, \s(a)) .
$$
An anti-automorphism $\s$ is an {\it anti-involution} if $\s^2=1$.

An important example of an anti-involution of Cend$_N$ is:
$$
\sigma (A(\partial,x) )= A^t(\partial, -x-\p)\tag 4.1
$$
where the superscript $t$ stands for the transpose of a matrix.

By Corollary 3.7 we know that all irreducible finite Cend$_N$-modules are of the
 form $(\alpha \in \Bbb C)$:
$$
A(\p,x)_\la v(\p) = A(-\la, \la+\p + \alpha ) v(\la+\p).
$$
Hence, twisting one of these modules by an automorphism of Cend$_N$ gives again
 one of these modules, and we get the following

\proclaim{Theorem 4.2} All automorphisms of Cend$_N$ are of the form:
$$
A(\p,x) \longmapsto C(\p+x)A(\p, x+\alpha)C(x)^{-1}, \quad
$$
where $\alpha\in \Bbb C$ and $C(x)$ is a matrix with a non-zero constant
 determinant.
\endproclaim

This result can be generalized as follows.

\proclaim{Theorem 4.3} Let $P(x)\in $Mat$_N \Bbb C[x]$  with det$\ P(x)\neq 0$.
 Then all automorphisms of Cend$_{N,P}$ are those that come from Cend$_N$ by
 restriction. More precisely, any automorphism is of the form:
$$
A(\p,x)P(x) \longmapsto C(\p+x)A(\p, x +\alpha )B(x)P(x), \tag 4.4
$$
where $\alpha\in\Bbb C$, and  $B(x)$ and $C(x)$ are invertible matrices in
 Mat$_N \Bbb C[x]$ such that
$$
P(x +\alpha )=B(x) P(x) C(x).\tag 4.5
$$
\endproclaim

\demo{Proof} Let $\pi'(a)=\pi (s(a))$, where $\pi$ is the
standard representation and $s$ is an automorphism of
Cend$_{N,P}$. Since it is equivalent to the standard
representation due to Corollary 3.7, we deduce that
$s(a(\p,x))=C(\p +x) a(\p,x +\alpha ) C(x)^{-1}$ for some
invertible (in Mat$_N \CC [x]$) matrix $C (x)$. But $ C(\p +x) $Cend$_{N,P}
C(x)^{-1}=$Cend$_{N,P}$ if and only if (4.5) holds. Indeed, we
have:  $C (\partial + x)$ $ P(x+\alpha ) C(x)^{-1}= A(\p ,x) P(x)$
for some $A(\p ,x) \in \Cend_N$.  Taking determinants of both
sides of this equality, we see that $\det A (\partial ,x)$ is a
non-zero constant. Hence $B(x):= P(x+\alpha ) C(x)^{-1}
P(x)^{-1}$ is invertible in Mat$_N \Bbb C[x]$, finishing the
proof.  \qed
\enddemo

\proclaim{Theorem 4.6} Let $P(x)\in $Mat$_N \Bbb C[x]$  with det$\ P(x)\neq 0$.
Then we have,

\noindent a) 
All non-zero  homomorphisms from Cend$_{N,P}$ to 
 Cend$_N$  are of the form:
$$
a(\p,x)P(x) \longmapsto S(\p+x)a(\p, x +\alpha )R(x), 
\tag 4.7 
$$
where $\alpha\in\Bbb C$,  and $R(x)$ and
$S(x)$ are matrices in
 Mat$_N \Bbb C[x]$ such that
$$
P(x +\alpha )=R(x) S(x).\tag 4.8
$$

\vskip .1cm

\noindent (b) All non-trivial anti-homomorphisms from Cend$_{N,P}$ to Cend$_N$ are of the
form: 
$$
a(\p,x) P(x) \longmapsto A(\p+x)a^t(\p, -\p - x +\alpha )B(x), 
\tag 4.9 
$$
where $\alpha\in\Bbb C$,  and $A(x)$ and
$B(x)$ are matrices in
 Mat$_N \Bbb C[x]$ such that
$$
P^t(-x +\alpha )=B(x) A(x).\tag 4.10
$$

\vskip .1cm

\noindent (c)  The  conformal algebra Cend$_{N,P}$ has an anti-automorphism (i.e. it is isomorphic to its opposite conformal algebra) if and only if the matrices $P^t(-x+\alpha)$ and $P(x)$ have the same elementary divisors for some $\alpha\in\CC$. In this case, 
all anti-automorphisms of Cend$_{N,P}$  are of the
form: 
$$
a(\p,x) P(x) \longmapsto Y(\p+x)a^t(\p, -\p - x +\alpha )W(x)P(x), 
\tag 4.11
$$
where  $Y(x)$ and
$W(x)$ are invertible matrices in
 Mat$_N \Bbb C[x]$ such that
$$
P^t(-x +\alpha )=W(x) P(x) Y(x).\tag 4.12
$$

\vskip .1cm

\noindent (d) The conformal  algebra Cend$_{N,P}$  has an
 anti-involution if and only if there exist an invertible in  Mat$_N \Bbb C[x]$ matrix $Y(x)$ such
 that  
 $$
 Y^t(-x + \alpha)P^t(-x+\alpha)=\ep P(x)Y(x)\tag 4.13
 $$
for $\ep=1$ or $-1$. In this case all anti-involutions are given by 
$$
\sigma_{P,Y,\ep ,\alpha}(a(\p,x)P(x))=\varepsilon Y(\p+x)a^t(\p,-\p-x+\alpha)Y^t(-x+\alpha)^{-1}P(x),\tag 4.14
$$
where $Y(x)$ is  an invertible in  Mat$_N \Bbb C[x]$ matrix satisfying  (4.13).
\endproclaim

\demo{Proof} a) Follows by the end of proof of Theorem 3.10(b).

\vskip .1cm

b)  Since composition of two anti-homomorphisms
is a 
homomorphism, using the anti-involution (4.1) we see that any
anti-homomorphism must be of the form 
$$
a(\p,x)P(x)\to R^t(-\p-x) a^t(\p, -\p-x+\alpha) S^t(-x)\tag 4.15
$$
with $P(x+\alpha)=R(x)S(x)$. Then, (4.9) and (4.10) follows by taking $A(x)=S^t(-x)$ and $B(x)=R^t(-\p-x)$.

\vskip .1cm

c) Let $\phi$ be an anti-automorphism of Cend$_{N,P}$. In particular, it is an anti-homomorphism
as in part b), whose image is Cend$_{N,P}$. Then, for all
$a(\p,x)P(x)\in$ Cend$_{N,P}$, we have that  $\phi(a(\p,x)p(x))=A(\p+x) a^t(\p,-\p-x+\alpha)B(x)\in$ Cend$_{N,P}$. Then taking $a(\p,x)$ the identity matrix we have that
$$
A(\p+x)B(x)= b(\p,x)P(x),\qquad\qquad \hbox{ for some }b(\p,x)\in\hbox{Cend}_{N,P}.\tag 4.16
$$
Recall that $P^t(-x+\alpha)=B(x)A(x)$. Taking determinant of  both sides of (4.16), and comparing its highest degrees in $x$, we deduce that $\det b(\p,x)$
and $\det A(x)$ are both (non-zero) constants. Now, from (4.16), we see that $A^{-1}(\p+x)b(\p,x)$ does not depend on $\p$. 
Then we have  $B(x)=W(x)P(x)$, where $W(x)=A^{-1}(\p+x)b(\p,x)$ is an invertible matrix. Therefore,
$$
\phi(a(\p,x)P(x))=A(\p+x)a^t(\p,-\p-x+\alpha)W(x)P(x),\tag 4.17
$$
with $A,W$ invertible matrices such that 
$$
W(x)P(x)A(x)=P^t(-x+\alpha).\tag 4.18
$$

\vskip .1cm

d) Now suppose that $\phi$ is an anti-involution. Then it is as in (4.11), and it also satisfies 
$\phi^2=id$. This condition implies that
$$
a(\p,x)P(x)=Y(\p+x)W^t(-\p-x+\alpha) a(\p,x)Y^t(-x+\alpha)W(x)P(x)\tag 4.19
$$
for all $a(\p,x)\in \hbox{
Cend}_{N,P}$. Denote $Z(x)=Y^t(-x+\alpha)W(x)$. Taking $a(\p,x)=$Id in (4.19) and using that $\det P(x)\neq 0$, , 
we have 
$Y(\p+x)W^t(-\p-x+\alpha)=Z^{-1}(x)$. Now, (4.19) becomes $a(\p,x)P(x)=Z^{-1}(x)a(\p,x)Z(x)P(x)$. Hence, we obtain  $Z(x)=\varepsilon\,Id$, with  $\varepsilon= 1$ or $-1$.
Thus, $Y^{-1}(X)=\varepsilon W^t(-x+\alpha)$. From (4.12) we deduce that 
$$
P(x)Y(x)=\varepsilon (P(-x+\alpha)Y(-x+\alpha))^t.\tag 4.20
$$
This condition is also sufficient. There exists an anti-involution if (4.20) holds for some invertible matrix $Y$, and it is given by
$$
\phi(a(\p,x)P(x))=\varepsilon Y(\p+x)a^t(\p,-\p-x+\alpha)Y^t(-x+\alpha)^{-1}P(x).\ \qed
$$
\enddemo

Two anti-involutions $\sigma,\tau$ of an associative conformal algebra $R$  are called {\it conjugate} if $\sigma =\varphi \circ \tau\circ \varphi^{-1}$ for some automorphism $\varphi$ of $R$. Recall that two matrices $a$ and $b$ in Mat$_N\CC[x]$ are called {\it $\alpha$-congruent} if
 $b=c^* a c$ for some invertible in Mat$_N\CC[x]$ matrix $c$, where $c(x)^*:=c(-x+\alpha)^t$. We shall simply call them {\it congruent} if $\alpha=0$. The following proposition gives us a characterization of equivalent anti-involutions $\sigma_{P,Y,\ep,\alpha}$ in Cend$_{N,P}$ (defined in (4.14)) and relates anti-involutions for different $P$.
 
\proclaim{Proposition 4.21} (a) The anti-involutions $\sigma_{P,Y_1,\ep_1,\alpha}$ and $\sigma_{P,Y_2,\ep_2,\gamma}$ of Cend$_{N,P}$ are conjugate if and only if $\ep_1=\ep_2$ and $P(x+\frac{\gamma-\alpha}{2}) Y_2(x+\frac{\gamma-\alpha}{2})$ is $\alpha$-congruent to $P(x)Y_1(x)$.

\vskip .2cm

\noindent (b) Let $\varphi_Y$ be the automorphism of Cend$_N$ given by $$
\varphi_Y(a(\p,x))=Y(\p+x)^{-1}a(\p,x)Y(x),
$$
where $Y$ is an invertible matrix in Mat$_N\CC [x]$, and let $P$ and $Y$ satisfying (4.13). Then 
$$
\sigma_{P,Y,\ep,\alpha}=\varphi_Y^{-1}\circ \sigma_{PY,I,\ep,\alpha}\circ \varphi_Y.\tag 4.22
$$

\vskip .2cm

\noindent (c) Let $c_\alpha$ be the automorphism of Cend$_N$ given by $c_\alpha(a(\p,x))=a(\p,x+\alpha)$, where $\alpha\in\CC$. Suppose that $P^t(-x+\alpha)=\ep P(x)$, for $\ep=1$ or $-1$, then   $Q(x):=P(x+\frac{\alpha}2)$ satisfies $Q^t(-x)=\ep Q(x)$ and 
$$
\sigma_{P,I,\ep,\alpha}=c_{\frac{\alpha}2}^{-1}\circ \sigma_{Q,I,\ep,0}\circ  c_{\frac{\alpha}2}.\tag 4.23
$$
\endproclaim

\demo{Proof} (a) Let $\varphi_{B,C, \alpha}$ be the automorphism of Cend$_{N,P}$ given by in (4.4) and (4.5). A straightforward computation shows that  $\varphi^{-1}_{B,C, \beta}\circ \sigma_{P,Y,\ep,\alpha} \circ \varphi_{B,C,\beta} = \sigma_{P,\bar Y,\ep,2\beta +\alpha}$, where $\bar Y(x)=C^{-1}(x-\beta) Y(x-\beta) B^t(-x+\alpha+\beta)$ and $P(x+\beta)=B(x)P(x)C(x)$. Hence, if  $\sigma_{P,Y_1,\ep_1,\alpha}$ and $\sigma_{P,Y_2,\ep_2,\gamma}$ are conjugate, then $\ep_1=\ep_2$ and $Y_2(x)=C^{-1}(x-\beta) Y(x-\beta) B^t(-x+\alpha+\beta)$, with $\beta=\gamma-\alpha/2$. Therefore, $P(x+\beta)Y_2(x+\beta)=B(x) P(x)Y_1(x) B^t(-x+\alpha)$, that is $P(x+\frac{\gamma-\alpha}{2}) Y_2(x+\frac{\gamma-\alpha}{2})$ is $\alpha$-congruent to $P(x)Y_1(x)$. 

Conversely, suppose that $P(x+\frac{\gamma-\alpha}{2}) Y_2(x+\frac{\gamma-\alpha}{2}) =  B(x) P(x)Y_1(x) B^t(-x+\alpha)$ for some $B(x)$  invertible matrix in Mat$_N\CC[x]$. Recall that $Y_1 $ and $Y_2$ are invertibles. Then $C(x):= Y_1(x) B^t(-x+\alpha) Y_2(x+ \frac{\gamma-\alpha}{2})^{-1}$ is an invertible matrix in Mat$_N\CC[x]$,  satisfies $P(x+ \frac{\gamma-\alpha}{2})= B(x) P(x)C(x)$, and it is easy to check that the anti-involutions are conjugated by the automorphism $\varphi_{B,C,\frac{\gamma-\alpha}{2}}$, proving (a).
 Part (b) and (c) are straightforward computations.   
\qed
\enddemo

\proclaim{Theorem 4.24} Any anti-involution of Cend$_N$ is, up to conjugation by an automorphism of Cend$_N$:
$$
a(\p, x)\mapsto a^*(\p , -\p -x),
$$
where  $^*$ is the adjoint with respect
to a non-degenerate symmetric or skew-symmetric
bilinear form over $\CC$.
\endproclaim

\demo{Proof} Using Theorem 4.6(d), we have that any anti-involution of Cend$_N$ has the form $\sigma(a(\p,x))=c(\p +x)a(\p , -
\p-x+\alpha)^t c(x)^{-1}$, where $c(x)$ is an invertible matrix such that $c(x)^t=\varepsilon c(-x+\alpha)$, with $\varepsilon=1$ or $-1$. By Proposition 4.21(c), we may suppose that $\alpha=0$. Now, the proof follows because $c(x)$ is congruent to a
constant symmetric or skew-symmetric matrix, by the following general
theorem of Djokovic. 

\enddemo

\proclaim{Theorem 4.25} (Djokovic, [D1-2]) If $A$ is invertible
in Mat$_N(\Bbb C[x])$ and $A^*=A$ (resp. $A^*=-A$) where
$A(x)^*=A(-x)$, then $A$ is congruent to a symmetric
(resp.~skew-symmetric) matrix over $\Bbb C$.
\endproclaim

\demo{Proof} The symmetric case follows by Proposition 5 in
[D1]. The skew-symmetric case was communicated to us by
D. Djokovic and we will give the details here. Suppose $A^*=-A$.
By Theorem (2.2.1), Ch.~7 in [Kn] it follows that $A$ has to be
isotropic, i.e. there exists a non-zero vector $v$ in $\Bbb
C[x]^N$ such that $v^* A v = 0$.  We can assume that $v$ is
primitive (i.e.,~the greatest common divisor of its coordinates is 1). But then
 $\Bbb C[x] v $ is a direct summand:  $\Bbb C[x]^N=\Bbb C[x] v\oplus M$, for
 some $\Bbb C[x]$-submodule $M$ of $\Bbb C[x]^N$. Then we have $\Bbb C[x]^N=
 (\Bbb C[x] v)^\perp\oplus M^\perp $ and $M^\perp$ is a free rank one $\Bbb
 C[x]$-module, that is $M^\perp = \Bbb C[x] w$ for some $w\in \Bbb C[x]^N$.
 Since $\Bbb C[x]v\subseteq (\Bbb C[x] v)^\perp$, the submodule $P=\Bbb C[x] v
 +\Bbb C[x] w$ is free of rank two. If $Q= M\cap (\Bbb C[x] v)^\perp$, then
 since $\Bbb C[x] v \subseteq (\Bbb C[x] v)^\perp$ we have $(\Bbb C[x]
 v)^\perp=\Bbb C[x] v \oplus Q$ and
$$
\Bbb C[x]^N=(\Bbb C[x]v)^\perp \oplus \Bbb C[x] w = P\oplus Q.
$$
Since $Q=P^\perp$,  the submodule generated by $v$ and
$w$ is a direct summand. Choose $w'\in P$ such that
 $v^* A w' = 1$. Then $v, w'$ must be a free basis of $P$ and the corresponding
 $2 \times 2$ block is of the form
$$
\pmatrix  0 & 1\\
          -1 & f
\endpmatrix
$$
for some skew element $f = g - g^*$ (cf. Proposition 5 [D1]). One can now
 replace $f$ by $0$, by taking the basis $v, w'-gv$, and use induction to finish
 the proof. \qed
\enddemo

\remark{Remark 4.26} We do not know any counter-examples to  the following generalization of Djokovic's
theorem: If $A\in\ $Mat$_N(\Bbb C[x])$ and $A^*=A$ (resp. $A^*=-A$) where
$A(x)^*=A(-x)$, then $A$ is congruent to  
a direct sum  of $1\times 1$ matrices of the form $(p(x))$ where $p$ is an even  (resp. odd) polynomial
and $2\times 2$ matrices of the form
$$
\pmatrix 0 &  q(x)\\ 
                -q(x) & 0 
\endpmatrix
$$
where $q(x)$ is an odd (resp. even) polynomial.
\endremark

\vskip .2cm

As a consequence of Theorem 4.6, we have the following result.

\proclaim{Theorem 4.27} Let $P(x) , Q(x)\in  $Mat$_N \CC[x]$ be two non-degenerate matrices.  Then Cend$_{N,P}$ is isomorphic to Cend$_{N,Q}$ if and only if there exist $\alpha\in\CC$ such that  $Q(x)$ and $P(x+\alpha)$ have the same elementary divisors.
 \endproclaim

\demo{Proof}   We may assume that $P$
is diagonal. Let $\phi:\ $Cend$_{N,P}\longrightarrow$Cend$_{N,Q}$ be an isomorphism.  In particular it is 
a homomorphism from Cend$_{N,P}$ to Cend$_{N}$
whose image is Cend$_{N,Q}$. Then, by Theorem 4.6(a), we have that 
  $\phi(a(\p,X)P(X))=A(\p+x) a(\p,x+\alpha)B(x)$, with  $P(x+\alpha)=B(x)A(x)$. In particular  
  $$
 A(\p+x) a(\p,x+\alpha)B(x) = Q(x)\tag 4.28
 $$ for some
$a(\p,x)P(x)\in$ Cend$_{N,P}$.

Taking determinant in both sides of (4.28), and comparing its highest degrees in $\p$, we can deduce that $\det A(x)$
 is constant. Now, define the isomorphism $\phi_2=\chi_A\circ\phi:\ $Cend$_{N,P}\to$Cend$_{N,QA}$, where
 $\chi_A(a(\p,x))=A^{-1}(\p+x)a(\p,x)A(x)$. Hence $\phi_2(a(\p,x)P(x))=a(\p,x+\alpha)B(x)A(x)$. Since $\phi_2$ is
 an isomorphism, we have that
 $$
 B(x)A(x)=D(x)Q(x)A(x)\qquad\hbox{ and }\qquad C(x)B(x)A(x)=Q(x)A(x)
 $$
 for some $C(x)$ and $D(x)$ (obviously $C$ and $D$  does not depend on $\p$).
  Comparing these two formulas, we have that $C(x)D(x)$ $=Id$. Then both are
 invertible matrices, and  $Q(x)A(x)=C(x)B(x)A(x)=C(x)P(x+\alpha)$ for some invertible  matrices $A$ and $C$.    \qed
\enddemo

\vskip .3cm


\head 5. On irreducible subalgebras of Cend$_N$
\endhead

\vskip .3cm

In this section we study the conformal analog of the Burnside Theorem. A
subalgebra of Cend$_N$ is called irreducible if it acts irreducibly on $\Bbb
C[\p]^N$. The following is the conjecture from [K2] on the classification of
such  subalgebras:

\proclaim{Conjecture 5.1} Any irreducible subalgebra of Cend$_N$ is either
 Cend$_{N,P}$ with det$\, P(x)\neq 0$ or $C(x+\p)\ \hbox{Cur}_N \ C(x)^{-1}$
 (i.e. is a conjugate of Cur$_N$), where det$\, C(x)$=1.
\endproclaim

The classification of  finite irreducible subalgebras follows from the
classification in [DK] at  the Lie algebra level:

\proclaim{Theorem 5.2} Any finite irreducible subalgebra of Cend$_N$ is a
 conjugate of Cur$_N$.
\endproclaim

\demo{Proof} Let $R$ be a finite irreducible subalgebra of Cend$_N$. Then the
Lie conformal algebra $R_-$ (with the bracket $[a_\lambda b]=a_\lambda
b-b_{-\partial-\lambda}a$), of course, still acts irreducibly on $\Bbb
C[\partial]^N$. By the conformal analogue of the Cartan-Jacobson theorem [DK]
applied to $R_-$, a conjugate $R_1$
of $R$ either contains the element $xI$ , or is contained in Mat$_{N}\Bbb
C[\partial ]$. The first case is ruled out since then $R_1$ is infinite. In the
second case, by the same theorem, $R_1$ contains Cur$\,\frak g$, where $\frak
g\subset$Mat$_N\Bbb C$ is a simple Lie algebra acting irreducibly on $\Bbb C^N$,
provided that $N>1$.

By the classical Burnside theorem, we conclude that $R_1=$Mat$_N\Bbb
C[\partial]$ in the case $N>1$. It is immediate to see that the same is true if
$N=1$ (or we may apply Theorem 2.2). \qed
\enddemo

\proclaim{Theorem 5.3} If $S\subseteq $Cend$_N$ is an irreducible
subalgebra such that $S$ contains the identity matrix~Id,
then $S=Cur_N$ or $S=$Cend$_N$.
\endproclaim

\demo{Proof} Since $Id\in S$, and using the idea of (1.7), we have that $S=\Bbb
 C[\p] A$,  where $A=S\cap $Mat$_N \Bbb C[x]$. Observe that $A$ is a subalgebra
 of Mat$_N \Bbb C[x]$. Indeed,
$$
P(x) Q(x)= P(x)_\lambda Q(x)_{|\lambda = -\p}\in S\qquad \hbox{ for all } P,Q\in
 A.
$$
In order to finish the proof, we should show that $A=$Mat$_N \Bbb
C$ or $A=$Mat$_N \Bbb C[x]$.  Observe that $A$ is invariant with respect to
 $d/dx$.  Because $P(x)_\lambda (Id)=P(\lambda+\p+x)\in \break\Bbb C[\lambda]\otimes
 S$.

Let $A_0 \subset \Mat_N \CC$ be the set of leading coefficients
of matrices from $A$.  This is obviously a subalgebra of $\Mat_N
\CC$ that acts irreducibly on $\CC^N$.  Otherwise we would have a
non-trivial $A_0$-invariant subspace $u \subset \CC^N$. Let $U$
denote the space of vectors in $\CC [\partial]^N$ whose leading
coefficients lie in $u$; this is a $\CC [\partial]$-submodule.
But  we have:
$$
a(x)_{\lambda} u (\partial) = a (\lambda + \partial) u
        (\lambda + \partial) = \sum_{j \geq 0}
        \frac{\lambda^j}{j!}
        (a (\lambda + \partial)u (\lambda + \partial))^{(j)}|_{\lambda=0} \, ,
$$
where $(j)$ stands for $j$-th derivative.
 Since both $A$ and $U$ are invariant with respect to
the derivative by the indeterminate, we conclude that $U$ is
invariant with respect to $A$, hence with respect to $S = \CC
[\partial]A$.

Thus, $A_0 = \Mat_N \CC$.  Therefore $A$ is a subalgebra of
$\Mat_N \CC [x]$ that contains $\Mat_N \CC$ and is
$d/dx$-invariant.  If $A$ is larger than $\Mat_N \CC$, applying
$d/dx$ a suitable number of times, we get that
$A$ contains a matrix of the form $xa$, where $a$ is a non-zero
constant matrix (we can always subtract the constant term).
Hence $A \supset x (\Mat_N \CC)a (\Mat_N \CC) = x\Mat_N\Bbb C$,
hence $A$ contains  $x^k \Mat_N(\Bbb C)$ for all $k \in \ZZ_+$.  \qed
\enddemo

\vskip .3cm


\head 6.  Lie conformal algebras $gc_N$, $oc_{N,P}$ and $spc_{N,P}$
\endhead

\vskip .3cm

A {\it Lie conformal algebra} $R$ is   a $\cp$-module endowed with a $\Bbb
 C$-linear map\break$
R\otimes R  \longrightarrow \Bbb C[\la]\otimes R $, $
a\otimes b  \mapsto [a_\la b]$, called the $\la$-bracket,
satisfying the following axioms $(a, b, c\in R)$,

\vskip .2cm

\noindent$(C1)_\la \qquad  [(\p a)_\la b]=-\la[ a_\la b],\qquad  [a_\la (\p
 b)]=(\la+\p) [a_\la b]$

\vskip .2cm

\noindent $(C2)_\la\qquad [a_\la b]=-[a_{-\p-\la} b]$

\vskip .2cm

\noindent $(C3)_\la\qquad [a_\la[b_\mu c] =[[a_\la b]_{\la+\mu} c] + [b_\mu
 [a_\la c]]$.

\

A {\it module} M over a conformal algebra  $R$ is  a $\cp$-module endowed with a
 $\Bbb C$-linear map  $R\otimes M \longrightarrow \Bbb C[\la]\otimes M$,
 $a\otimes v  \mapsto a_\la v$,
satisfying the following axioms $(a,\, b \in R),\ v\in M$,

\vskip .2cm

\noindent$(M1)_\la \qquad   (\p a)_\la^M v= [\p^M, a_\la^M]v=-\la a_\la^Mv   ,$

\vskip .2cm

\noindent $(M2)_\la\qquad [a_\la^M, b_\mu^M]v=[a_{ \la} b]_{\la+\mu}^Mv$.

\vskip .2cm

Let $U$ and $V$ be  modules over a conformal algebra $R$. Then , the
 $\cp$-module $N:=$ Chom$(U,V)$ has an $R$-module structure defined by
$$
(a_\la^N\varphi)_\mu u=a_\la^V(\varphi_{\mu-\la}u)-\varphi_{\mu-\la}(a_\la^U u),\tag 6.1
$$
where $a\in R$, $\varphi\in N$ and $u\in U$. Therefore, one can define the
 contragradient conformal $R$-module $U^*=$Chom$(U,\Bbb C)$, where $\Bbb C$ is
 viewed as the trivial $R$-module and $\cp$-module. We also define the tensor
 product $U\otimes V$  of $R$-modules as the ordinary tensor product with
 $\cp$-module structure $(u\in U, v\in V)$:
$$
\p(u\otimes v)\,=\, \p u\otimes v + u\otimes \p v
$$
and $\la$-action defined by $(r\in R)$:
$$
r_\la (u\otimes v)\,=\, r_\la u\otimes v + u\otimes r_\la v.
$$

\proclaim{Proposition 6.2} Let $U$ and $V$ be two $R$-modules. Suppose that $U$ has  finite rank as a $\cp$-module.
 Then $U^*\otimes V\simeq $ Chom$(U,V)$ as $R$-modules, with the identification
 $(f\otimes v)_\la (u)=f_{\la+\p^V} (u) \, v$, $f\in U^*, \, u\in U$ and $v\in
 V$.
\endproclaim

\demo{Proof} Define $\varphi: U^*\otimes V\to $Chom$(U,V)$  by
 $\varphi(f\otimes v)_\la (u)=f_{\la+\p^V} (u)\, v$. Observe that  $\varphi$ is
 $\cp$-linear, since
$$
\aligned
\varphi(\p(f\otimes v))_\la (u) &=\varphi( \p f\otimes v+ f\otimes \p v)_\la
 (u)=(\p f)_{\la+\p^V} (u) \, v + f_{\la+\p^V} (u) \, \p v\\
&= -(\la+\p^V) f_{\la+\p^V} (u) \, v + f_{\la+\p^V} (u) \, \p v = -\la
 f_{\la+\p^V} (u) \, v \\
&= -\la \varphi(f\otimes v)_\la (u) = \p(\varphi(f\otimes v))_\la (u)
\endaligned
$$
and $\varphi$ is a homomorphism, since
$$
\aligned
\varphi\Bigl( r_\la(f\otimes v)\Bigr)_\mu (u)&= \varphi \Bigl( r_\la f\otimes v
 + f \otimes r_\la v \Bigr)_\mu (u)\\
&= (r_\la f)_{\mu+\p^V} (u) \, v   +  f_{\mu+\p^V} (u) \, (r_\la v)\\
&= - f_{\mu-\la+\p^V} (r_\la u) \, v   +  f_{\mu+\p^V} (u) \, (r_\la v)
\endaligned
$$
and
$$
\aligned
\Bigl(r_\la\bigl( \varphi(f\otimes v)\bigr)\Bigr)_\mu (u)&=   r_\la
 \Bigl(\varphi (f\otimes v) _{\mu-\la} (u) \Bigr) - \varphi(f \otimes v
 )_{\mu-\la} (r_\la u)\\
&= r_\la (f_{\mu-\la +\p^V} (u) \, v  ) -  f_{\mu -\la +\p^V} (r_\la u) \, v \\
&=  f_{\mu+\p^V} (u) \, (r_\la v) - f_{\mu-\la+\p^V} (r_\la u) \, v .
\endaligned
$$
The homomorphism $\varphi$ is always injective. Indeed, if  $\varphi(f\otimes v)=0$, then $f_{\mu+\p^V} (u) v=0$ for all $u\in U$. Suppose that $v\neq 0$, then $f_{\la +\p^V}=0$, that is $f=0$.

It remains to prove that $\varphi$ is surjective provided that $U$  has
 finite rank as a $\cp$-module. Let $g\in \hbox{Chom}(U,V)$, and $U=\CC[\p]\{u_1,\cdots , u_n\}$. Then, there exist $v_{ik}\in U$ such that
$$
g_\la(u_i)=\sum_{k=0}^{m_i} (\la+\p^V)^k v_{ik}= \sum_{k=0}^{m_i} \varphi(f_{ik}\otimes v_{ik})_\la (u_i),
$$
where $f_{ik}\in U^*$ is defined (on generators) by $f_{ik}(u_j)= \delta_{i,j} \la^k$. Therefore, $g=\varphi (\sum_{i=0}^n\sum_{k=0}^{m_i} f_{ik}\otimes v_{ik})$, finishing the proof. \qed\enddemo

\vskip .2cm

In general, given any associative conformal algebra $R$ with $\la$-product
 $a_\la b$, the $\la$-bracket defined by
$$
[a_\la b]:=a_\la b-b_{-\p-\la} a \tag 6.3
$$
makes $R$ a Lie conformal algebra.

Let $V$ be a finite $\cp$-module. The $\la$-bracket (6.3) on Cend$\,V$, makes it
 a Lie conformal algebra denoted by gc$\,V$ and called the {\it general
 conformal algebra} (see [DK], [K1] and [K2]). For any positive integer $N$, we define gc$_N:=$gc$\, \Bbb
 C[\p]^N=Mat_N \Bbb C[\p,x]$, and the $\la$-bracket
 (6.3) is by (1.1):
$$
[A(\partial,x)_\lambda B(\partial,x)]=A(-\lambda,x+\lambda+\partial)
 B(\lambda+\partial,x)-  B(\lambda+\partial,-\la+x)A(-\lambda,x ).
$$

Recall that, by Theorem 4.24, any anti-involution in Cend$_N$ is, up to conjugation
$$
\sigma_*(A(\p,x))=A^*(\p,-\p-x),\tag 6.4
$$
where $*$ stands for the adjoint  with respect to a non-degenerate
 symmetric or skew-symmetric bilinear form over $\CC$. These anti-involutions give us two important subalgebras of $gc_N$: the set of $-\sigma_*$
fixed points  is the {\it orthogonal conformal algebra}
$oc_N$ (resp. the {\it  symplectic conformal algebra} $spc_N$), in the symmetric
(resp. skew-symmetric) case.

\vskip .2cm

\proclaim {Proposition 6.5} The subalgebras $oc_{N}$ and $spc_{N}$ are 
simple.\endproclaim

\demo{Proof} We will prove that $oc_{N}$ is simple. The proof  for $spc_N$ is  similar. Let $I$ be  a non-zero ideal of $oc_{N}$. Let  $0\neq A(\p,x)\in I$, then $A(\p, x)=\sum_{i=0}^m \p^i a_i(x)=\sum_{j=0}^n \p^j \tilde a_j(\p+x)$, with $a_i(x), \tilde a_j(x)\in \, $Mat$_N\CC[x]$. Now, using that $A(\p,x)=-A^t(\p,-\p-x)$, we obtain that $n=m$ and $a_i(x)=-\tilde a_i^t(-x)$.  Computing the $\lambda$-bracket 
$$
[x E_{ij}-(-\p-x) E_{ji}\,_\lambda \, A(\p,x)]= \la^{m+1} (E_{ij}a_m(x)-a_m^t(-\p-x)E_{ji})+ \la^m \dots 
$$
we deduce that $E_{ij}a_m(x)-a_m^t(-\p-x)E_{ji}\in I$, with $a_m\neq 0$. By  taking appropriate $i$ and $j$, we have that there exist polynomials $b_k(x)$ such that \break$\sum_{k=1}^N (b_k(x)E_{ik} -b_k(-\p-x)E_{ki})\in I$, with $b_r\neq 0$ for some $r\neq i$. Now by computing  
$
[(2x+\p) E_{rr}\,_\lambda \, \sum_{k=1}^N (b_k(x)E_{ik} -b_k(-\p-x)E_{ki})]
$ 
and looking at its leading coefficient in $\lambda$, we show that $E_{ri}-E_{ir}\in I$, with $r\neq i$. Taking brackets with elements in $o_N$, we have  $E_{jl}-E_{lj}\in I$ for all $j\neq l$. Now,  we can see from the $\la$-brackets 
$[ x E_{ri}- (-\p-x) E_{ir}\,_\lambda\,E_{ir}-E_{ri}]= (2x+\p)(E_{ii}-E_{rr})$ and    $[(2x+\p)E_{ii}\,_\lambda\, (2x+\p)(E_{ii}-E_{rr})]= \lambda (2x+\p)E_{ii}$, 
that $(2x+\p)E_{ii}\in I$ for all $i$. The other generators are obtained by $(k\neq i,j)$
$$
[(-x)^k E_{ik}-(\p+x)^k E_{ki}\,_\lambda \, E_{jk}-E_{kj}]_{|_{\la=0}}= x^k E_{ij}-(-\p-x)^k E_{ji}.
$$
Similarly, we can see that $(x^k -(-\p-x)^k)E_{ii}\in I$, finishing the proof.
  \qed
\enddemo

The conformal subalgebras $oc_N$ and $spc_N$, as well as the anti-involutions given by  (6.4), and their generalizations  can be described in terms of conformal bilinear forms. Let $V$ be a $\CC[\p]$-module. A {\it conformal bilinear form} on $V$ is a $\CC$-bilinear map $\< \  , \  \>_\la : V\times V\to \CC[\la]$ such that
$$
\<\p v,w\>_\la = -\la \< v,w\>_\la = -\< v, \p w\>_\la, \, \  \hbox{ for all } v,w\in V.
$$
The conformal bilinear form is {\it non-degenerate} if $\< v,w \>_\la =0$ for all $w\in V$, implies $v=0$. The conformal bilinear form is {\it symmetric } (resp. {\it skew-symmetric}) if $\< v,w \>_\la= \ep  \< w , v\>_{-\la}$ for all $v,w\in V$, with $\ep=1$ (resp. $\ep=-1$).

\vskip .2cm 

Given a  conformal bilinear form on a $\CC[\p ]$-module $V$, we have a homomorphism of  $\CC[\p]$-modules,  $L : V\to V^*$, $v\mapsto L_v$, given as usual by
$$
(L_v)_\la w= \< v, w\>_\la ,  \quad v\in V.\tag 6.6
$$
Let  $V$ be a free finite rank $\CC[\p]$-module and  fix $\beta=\{e_1,\cdots , e_N\}$  a $\CC[\p]$-basis of $V$. Then  {\it the matrix  of $\<\ ,\ \>_\la$ with respect to $\beta$} is defined  as   $P_{i,j}(\la)=\< e_i,e_j\>_\la$. Hence, identifying $V$  with $\CC[\p]^N$, we have
$$
\< v(\p) , w(\p)\>_\la = v^t(-\la) P(\la) w(\la).\tag 6.7
$$
Observe  that $P^t(-x)= \ep P(x)$ with $\ep=1$ (resp. $\ep=-1$) if the conformal bilinear form is symmetric (resp. skewsymmetric). We also have that Im$\  L= P(-\p) V^*$, where $L$ is defined in (6.6). Indeed, given $v(\p)\in V$, consider $g_\la\in V^*$ defined by $g_\la(w(\p))= v^t(-\la) w(\la)$, then by (6.7)
$$
(L_{v(\p)})_\la w(\p)= v^t(-\la) P(\la) w(\la) =g_\la(P(\p)w(\p))=(P(-\p) g)_\la(w(\p)),
$$
where in the last equality we are identifying $V^*$ with $\CC[\p]^N$ in the natural way, that is $f\in V^*$ corresponds to $(f_{-\p} e_1 , \cdots , f_{-\p} e_N)\in \CC[\p]^N$.
Therefore, if the conformal bilinear form is non-degenerate, then $L$ gives an isomorphism between $V$ and $P(-\p) V^*$, with $\det P\neq 0$.

Suppose that we have a non-degenerate conformal bilinear form on $V=\CC[\p]^N$ which is also symmetric or skew-symmetric. Denote by $P(\la)$ the matrix of this  bilinear form with respect to the standard basis of $\CC[\p]^N$. 
Then for each $a\in\ $Cend$_N$ and  $w\in V$, the map $f^{a,w}_\lambda(v):=\langle w,a_\mu
v\rangle_{\lambda-\mu}$ is in $\Bbb C[\mu]\otimes V^*$, that is $f^{a,w}_\lambda$ is a
$\Bbb C$-linear map, $f^{a,w}_\lambda(\partial v)=\lambda f^{a,w}_\lambda(v)$ and depends
polynomialy on $\mu$, because $\deg_\mu f^{a,w}_\la(v)\leq \max \{\deg_\mu f^{a,w}_\la(e_i)\ : \, i=1,\cdots ,N\}$. Observe that if we restrict to Cend$_{N,P}$, then $f^{aP,w}_\lambda= (P(-\p)f^{a,w})_\lambda\in \ \hbox{Im}\ L$. Therefore, since $\langle\ ,\ \rangle_\lambda$ is
non-degenerate, there exists a unique $(aP)^*_\mu w\in \Bbb C[\mu]\otimes V$ such that
$f^{aP,w}_\lambda(v)=\langle w,aP_\mu v\rangle_{\lambda-\mu}=\langle (aP)^*_\mu w,v\rangle_{\lambda}$. Thus, we have
attached to each $aP\in\ $Cend$_{N,P}$ a map $(aP)^*:V\longrightarrow \Bbb C[\mu]\otimes V$, $w\mapsto (aP)^*_\mu w$, where
the vector $(aP)^*_\mu w$ is determined by the identity
$$
\langle aP_\mu v,w\rangle_{\lambda}=\langle v,(aP)^*_\mu w\rangle_{\lambda-\mu}.
$$
Observe that $(aP)^*_\mu(\partial w)=(\partial+\mu) \  (aP)^*_\mu \,w$, that is $(aP)^*\in\ $Cend$\,V$. Indeed,
$$
\aligned
\langle v,(aP)^*_\mu \,(\partial w)\rangle_{\lambda-\mu} & =\langle aP_\mu v, \partial
w\rangle_{\lambda}=\lambda\,\langle aP_\mu v, w\rangle_{\lambda}\\
& =-\langle \partial\, (aP_\mu\, v),  
w\rangle_{\lambda}=\langle \mu \, aP_\mu v,  w\rangle_{\lambda}-\langle aP_\mu \,\partial v, w\rangle_{\lambda}\\
& =\mu \,\langle   v,  (aP)^*_\mu w\rangle_{\lambda-\mu}\,-\,\langle \partial\, v,  (aP)^*_\mu w\rangle_{\lambda-\mu}\\
& =\langle v,  (\mu+\partial)\, (aP)^*_\mu w\rangle_{\lambda-\mu}.
\endaligned
$$
Moreover we have the following result:

\vskip .2cm

\proclaim{Proposition 6.8} (a) Let $\<\  ,\ \>_\la$ be a non-degenerate symmetric or skew-symmetric conformal bilinear form on $\CC[\p]^N$, and denote by $P(\la)$ the matrix of $\<\  ,\ \>_\la$ with respect to the standard basis of $\CC[\p]^N$ over $\CC [\p]$. Then the map   $aP\mapsto (aP)^*$ from Cend$_{N,P}$ to Cend$_N$ defined by
$$
\< a_\mu v, w\>_\la = \< v , a^*_\mu w\>_{\la-\mu}.\tag 6.9
$$
is the anti-involution of Cend$_{N,P}$ given by 
$$
(a(\p,x)P(x))^*= \ep a^t(\p,-\p-x)P(x),\tag{6.10}
$$ 
where  $P^t(-x)=\ep P(x)$ with $\ep=1$ or $-1$, depending on whether the conformal bilinear form   is symmetric or skew-symmetric.

\noindent (b) Consider the Lie conformal subalgebra of $gc_N$ defined by 
$$
\aligned g_* &=\{ a\in\hbox{Cend}_{N,P}\, :\, a^*=-a\,\}\\
&=
\{a\in\text{Cend}_{N,P}\,:\, \langle a_\mu v, w\rangle_\lambda + \langle v,a_\mu w\rangle_{\lambda-\mu}=0,\ \ \hbox{ for all } v,w\in\Bbb C[\partial]^N\},
\endaligned
$$
where $^*$ is defined by (6.10). Then under the pairing (6.6) we have $\CC[\p]^N\simeq P(-\p)(\CC[\p]^N)^*$ as $g_*$-modules.\endproclaim

\demo{Proof} (a) First let us check that $\varphi(aP)=(aP)^*$ defines an anti-homomorphism from Cend$_{N,P}$ to Cend$_N$. Since $(a,b\in\hbox{Cend}_{N,P})$ 
$$
\aligned 
\langle  v, (a_\mu b)^*_\gamma w\rangle_{\lambda-\gamma} & =\langle (a_\mu b)_\gamma v, w\rangle_{\lambda}=
\langle a_\mu(b_{\gamma-\mu} v),  w\rangle_{\lambda}\\
&= \langle b_{\gamma-\mu}  v, a^*_\mu w\rangle_{\lambda-\mu}=\langle v,  b^*_{\gamma-\mu} (a^*_\mu
w)\rangle_{\lambda-\gamma}\\
& =\langle  v, (b^*_{\gamma-\mu}a^*)_\gamma w\rangle_{\lambda-\gamma},
\endaligned
$$
we have that $\varphi(a_\mu
b)_\gamma=(\varphi(b)_{\gamma-\mu}\varphi(a))_\gamma=(\varphi(b)_{\partial-\mu}\varphi(a))_\gamma$. 

Now,
using Theorem 4.6(b), we have that 
$$
\varphi(a(\p,x)P(x))= A(\p+x) a^t(\p,-\p-x+\alpha)B(x),
$$
with $\alpha\in \CC$ and $P^t(-x+\alpha)=B(x)A(x)$. Replacing $\varphi(aP)$ in (6.9) and using (6.7), we obtain
$$
P(\la-\mu)a^t(-\mu, \mu-\la)P(\la)=P(\la-\mu)A(\la-\mu) a^t(-\mu, \mu-\la+\alpha) B(\la),  \hbox{ for all } a(\p,x).\tag 6.11
$$
Taking $a(\p,x)=I$ and using that $\det P\neq 0$,  we have $P(\la)=A(\la-\mu)B(\la)$. Since the left hand side does not depend on $\mu$, we get $A=A(x)\in$Mat$_N\CC$, with $\det A\neq 0$. Using that $\ep P(x-\alpha)=P^t(-x+\alpha)=B(x)A$, then (6.11) become
$$
a^t(-\mu, \mu-\la) \ep B(\la+\alpha)A=A a^t(-\mu, \mu-\la+\alpha) B(\la),\quad \hbox{ for all } a(\p,x).
$$
In particular, we have $\ep B(\la+\alpha)A=A  B(\la)$. Hence  
$
a^t(-\mu, \mu-\la) A=A a^t(-\mu, \mu-\la+\alpha)$  for all $a(\p,x)$, getting $\alpha=0$ and  $A=cI$. Therefore,
$$
\varphi(a(\p,x)P(x))= \ep a^t(\p,-\p-x)P(x),
$$
with  $P^t(-x)=\ep P(x)$ with $\ep=1$ or $-1$, depending on whether the conformal bilinear form   is symmetric or skew-symmetric, getting (a).

\noindent (b) Using  (6.6), we obtain for all $a\in g_*$ and $ v,w\in \CC[\p]^N$ that  
$$
(L_{a_\mu v})_\la (w)= \< a_\mu v , w\>_\la= -\< v ,a_\mu w\>_{\la-\mu}= -(L_v)_{\la-\mu}(a_\mu w)= (a_\mu(L_v))_\la(w).
$$
finishing the proof.\qed\enddemo

\vskip .2cm

Observe that oc$_N$ (resp. spc$_N$), can be described as the subalgebra $g_*$ of $gc_N$ in Theorem 6.8(b), with respect to the conformal bilinear form 
$$
\langle p(\partial)v,q(\partial)w\rangle_\lambda= p(-\lambda)q(\lambda) \, ( v,w )\qquad\text{ for all }
v,w\in\Bbb C^N,
$$
where $(\cdot,\cdot)$ is a non-degenerate symmetric (resp.  skew-symmetric) bilinear form on $\CC^N$. For general $P$, see (6.16) below.

Then, $oc_N$ (resp. $spc_N$) is the $\cp$-span of $\{y^n_A:=x^n A-(-\p-x)^n A^*\,:\, A\in
\hbox{Mat}_N\Bbb C \}$, where $*$ stands for the adjoint  with respect to a non-degenerate
 symmetric (resp. skew-symmetric) bilinear form over $\CC$. Therefore we have that $gc_N=oc_N\oplus M_N$ (resp. $gc_N=spc_N\oplus M_N$), where
$M_N$ is the set of $\s_*$-fixed points, i.e. $$ M_N=\cp\hbox{-span of
}\{w^n_A:=x^nA+(-\p-x)^nA^*\,:\, A\in \hbox{Mat}_N\Bbb C  \}.\tag 6.12
$$
We are using the same notation $M_N$ in the symmetric and skew-symmetric case.
Observe that $M_N$ is an $oc_N$-module (resp. $spc_N$-module) with the action given by
$$
\aligned
y^n_A\ _\la\ w^m_B &= (\la+\p+w_{AB})^n w^m_{AB}-(-\p-w_{A^*B})^n w^m_{A^*B}\\
&+(-1)^n(-\la -\p-w_{AB^*})^{m+n}-(-\la +w_{BA})^m w^n_{BA}
\endaligned\tag 6.13
$$

\vskip .2cm

Let us give a more conceptual understanding of the module $M_N$. Let $V=\cp^N$.
 By definition, $V^*=$Chom$(V,\Bbb C)=\{\alpha:\cp^N\to\Bbb C[\la]\,:\,
 \alpha_\la\p=\la\alpha_\la\}$ and given $\alpha\in V^*$  it is completely
 determined by  the values in the canonical basis $\{e_i\}$ of $\Bbb C^N$, this
 is $p_\alpha(\la) :=(\alpha_\la e_1,\cdots,\alpha_\la e_N) \in\Bbb C[\la]^N$.
 Thus, we may identify $V^*\simeq \Bbb C[\la]^N$ and $\Bbb C[\p]$-module
 structure given by $(\p p)(\la)=-\la p(\la)$.

We have that $gc_N$  acts on $V$ by the $\la$-action
$$
A(\p,x)\,_\la\, v(\p)=A(-\la, \la +\p) v(\la+\p), \quad v(\p)\in \Bbb C[\p]^N,
$$
and on $V^*$  by the contragradient action, given by
$$
A(\p,x)\,_\la\, v(\p)= -  \ ^tA(-\la, -\p) v(\la+\p), \quad v(\p)\in \Bbb
 C[\p]^N.
$$
It is easy to check that $(V^*)^*\simeq V$ as $gc_N$-modules. Observe that by Proposition 6.8(b),  $V\simeq V^*$ as
 $oc_N$-modules and $spc_N$-modules.

 We define the {\it 2nd exterior power} $\Lambda^2(V)$ and  the {\it 2nd
 symmetric power} $S^2(V)$ in the usual way with the induced $\cp$-module
 and $gc_N$-module structures.

\vskip .1cm

\proclaim{Proposition 6.14} (a) $V\otimes V= S^2(V)\oplus \Lambda^2(V)$ is the
 decomposition of $V\otimes V$ into a direct sum of irreducible $gc_N$-modules.
 And $V^*\otimes V$ is isomorphic to the adjoint representation of $gc_N$.

\noindent (b) $gc_N\simeq V\otimes V= S^2(V)\oplus \Lambda^2(V)$ is the
 decomposition of $gc_N$    into a direct sum of irreducible $oc_N$-modules, where 
  $\Lambda^2(V)$ is isomorphic to the adjoint representation of $oc_N$,
 and $M_N\simeq S^2(V)$ as $oc_N$-modules.

\noindent (c) $gc_N\simeq V\otimes V= S^2(V)\oplus \Lambda^2(V)$ is the
 decomposition of $gc_N$    into a direct sum of irreducible $spc_N$-modules, where 
  $S^2(V)$ is isomorphic to the adjoint representation of $spc_N$,
 and $M_N\simeq \Lambda^2(V)$ as $spc_N$-modules.
\endproclaim

\demo{Proof} (a) Follows from Proposition 6.2 and part (b).

\vskip .3cm

\noindent (b)  Define $\varphi:V\otimes V\to gc_N$ by
$$
\varphi(p(\p)e_i\otimes q(\p) e_j)= p(-x)q(x+\p)E_{ji}
$$
It is easy to check that this is an $oc_N$-module isomorphism. Note that $\s_*$
 defined in (6.4) corresponds via $\varphi$ to  $
\s\bigl( p(\p)e_i\otimes q(\p) e_j \bigr)= q(\p) e_j\otimes p(\p)e_i $.
 Therefore it is immediate that $M_N\simeq S^2(V)$ and  $\Lambda^2(V)\simeq
 oc_N$. It remains to see that $M_N$ is an irreducible $oc_N$-module. Let $W\neq
 0$ be a $oc_N$-submodule of $M_N$ and $0\neq
 w(\p,x)=\sum_{i,j}q_{ij}(\p,x)E_{ij}\in W$. We may suppose that $q_{11}\neq 0$.
 Computing $[y^1_{E_{11}}\, _\la w(\p,x)]$ and looking at the highest degree of
 $\la$ that appears in the component $E_{11}$, we deduce that there exists in
 $W$ an element of the form $w'=\sum_i (p_i(\p,x)E_{1i}+q_i(\p,x)E_{i1})$, with
 $p_1=q_1=1$. Now, computing $ [y^1_{E_{12}} \,_\la w'(\p,x)]$ we have that
 $w''=r(\p,x)E_{11}+ w^1_{E_{12}}+\hbox{ terms out of the first column and
 row}\in W$. And from $[y^1_{E_{11}}\, _\la w''(\p,x)]$ and looking at the
 highest degree in $\la$, we have that if $r(\p,x)$ is non constant,
 $w^0_{E_{11}}\in W$, and if  $r(\p,x)$ is  constant,
 $w^0_{E_{11}}+w^1_{E_{12}}\in W$. In both cases, by (6.6)
we have that $w^0_I\in W$. Now, looking at ($n>>0$ and $A$ arbitrary)
$$
y^n_A\,_\la w^0_I=\la^n 2w^0_A+\la^{n-1}2n(\p w^0_A +w^1_A)+
\la^{n-2}2\binom n 2 (\p^2 w^0_A +2\p w^1_A+w^2_A) + \cdots
$$
we get $W=M_N$, finishing part (b).

(c) The proof is similar to (b), with  $\varphi:V\otimes V\to gc_N$ defined by 
$
\varphi(p(\p)e_i\otimes q(\p) e_j)= p(-x)q(x+\p)E_{ij}^\dag$, where 
$E_{ij}^\dag= -E_{j ,\frac N 2 +i}$, $E_{\frac N 2 +i,\frac N 2 +j}^\dag= E_{\frac N 2 +j ,i}$, $E_{i,\frac N 2 +j}^\dag= -E_{\frac N 2 +j ,\frac N 2 +i}$ and $E_{\frac N 2 +i,j}^\dag= -E_{j ,i}$, for all $1\leq i , j\leq  \frac N 2$.  
\qed\enddemo

\vskip .5cm

Observe that $gc_{N,P}:= gc_N P(x)$ is a conformal subalgebra of $gc_N$, for
 any $P(x)\in \hbox{Mat}_N \Bbb C[x]$.

A matrix $Q(x)\in\ $Mat$_N\CC[x]$ will be called {\it hermitian} (resp. {\it skew-hermitian}) if
$$
Q^t(-x)=\varepsilon Q(x)\qquad  \text{ with     } \varepsilon=1 \quad (\text{ resp.   } \varepsilon=-1).
$$
Denote by $o_{P,Y,\varepsilon,\alpha}$ the subalgebra of gc$_{N,P}$ of 
$-\sigma_{P,Y,\varepsilon,\alpha}$-fixed points. By Proposition 4.21 (b)-(c), we have the following
isomorphisms, obtained by conjugating by  automorphisms of Cend$_N$
$$
o_{P,Y,\varepsilon,\alpha}\simeq o_{PY,I,\varepsilon,\alpha}\simeq o_{Q,I,\varepsilon,0}, \tag 6.15
$$
where $Q(x)=(PY)(x+\alpha/2)$ is hermitian or skew-hermitian, depending on whether  $\varepsilon=1$ or $-1$. Therefore, up to conjugacy,  we may
restrict our attention to the  family of subalgebras  (6.15), that is it suffices to consider the anti-involutions 
$$
\sigma_{P,I,\varepsilon,0}(a(\p,x)P(x))=\varepsilon a^t(\p,-\p-x) P(x)
$$
where $P$ is non-degenerate hermitian or skew-hermitian, depending on whether $\varepsilon = 1$ or $-1$. From now on we shall use the following notation
$$
\aligned
oc_{N,P}:= & o_{P,I,1,0}\qquad \text { if } P \text{ is hermitian }\\
spc_{N,P}:= & o_{P,I,-1,0}\qquad\text{ if } P \text{ is skew-hermitian}.
\endaligned\tag 6.16
$$
These subalgebras are those obtained in Theorem 6.8(b) in a more invariant form. In the special case $N=1$ and $P(x)=x$, the involution $\sigma_{x,I,-1,0}$ is the conformal version of the
involution given by Bloch in [B].

Note  that gc$_{N,P}\simeq oc_N\cdot P(x)\oplus M_N\cdot P(x)$. If $P$ is hermitian, then $oc_{N,P}=oc_N\cdot
P(x)$ and $M_N\cdot P(x)$ is an $oc_{N,P}$-module. If $P$ is skew-hermitian, then  $spc_{N,P}=M_N\cdot P(x)$, and $oc_N\cdot P(x)$ is a $spc_{N,P}$-module.

\vskip .2cm

\remark{Remark 6.17}   (a)  The subalgebras $gc_N$,  $gc_{N,x I}$,  $oc_N$  and   
$spc_{N,x I}$
 contain the  conformal
 Virasoro subalgebra $\CC[\p] (x+\alpha \p) I$, for $\alpha$ arbitrary, $\alpha=0$, $\alpha=\frac 1 2$ and $\alpha=0$ respectively. 

\noindent (b) Let $J=\pmatrix 0 & I\\ -I & 0 \endpmatrix$, then by (6.15) we obtain
$$
spc_N=o_{I,J,-1,0}\simeq o_{J,I,-1,0} =spc_{N,J}.
$$

\noindent (c) Using the proof of Proposition 6.5, one can prove that  $oc_{N , P }$ and $spc_{N , P }$,
with $\det P(x)\neq 0$, are simple  if $P(x)$ satisfies  the property that for each $i$ there exists $j$ such that  $\deg P_{ij}(x) > \deg P_{ik}(x)$ for all $k\neq j$. 
\endremark

\vskip .2cm 

\proclaim{Proposition 6.18} The subalgebras $oc_{N , P }$ and $spc_{N , P }$,
with $\det P(x)\neq 0$, acts irreducibly on $\Bbb C[\p]^N$.
\endproclaim

\demo{Proof}
Let  $M$ be a non-zero oc$_{N,P}$-submodule of $\CC[\p]^N$ and take
 $0\neq v(\p)\in M$. Since $\det P(x)\neq 0$, there exists $i$ such 
that $P(y)v(y)$ has  non-zero $i$th-coordinate that we shall denote
by $b(y)$. Recall that  $\{(x^k A-(-\p-x)^k A^t)P(x)\ |\ A\in \hbox{Mat}_N\CC\}$ generates $oc_{N,P}$. Now, looking at the highest degree in $\lambda$ in
$$
(2x+\p) E_{ii}P(x)\,_\lambda \,v(\p)=(\lambda+2\p)
 b(\p+\lambda)e_i
$$
we  deduce that $e_i\in M$. Now, since the $i$th-column of $P=(P_{r,j})$ is
non-zero, we can take $k$ such that $P_{k,i}(x)\neq 0$ has maximal degree in $x$, in the $i$th-column. Then,  considering the
$\lambda$ action of $(x E_{jk}-(-\p-x) E_{kj})P(x)$ on $e_i$, for
$j=1,\cdots, N$,
 and looking at the highest degree in $\lambda$, we  have that
 $e_j\in M$ for all $j=1,\cdots ,N$. Therefore $M=\Bbb C[\p]^N$. A similar
 argument also works for spc$_{N,P}$.  \qed
\enddemo

\proclaim{Proposition 6.19} (a) The subalgebras $oc_{N,P}$ and $oc_{N,Q}$ (resp. $spc_{N,P}$ and \break $spc_{N,Q}$) are conjugated by an automorphism of Cend$_N$ if and only if $P$ and $Q$ are congruent hermitian (resp. 
skew-hermitian) matrices. 

\noindent (b) The subalgebras $oc_{N,P}$ and $spc_{N,Q}$ are not conjugated  by any automorphism of Cend$_N$.
\endproclaim

\demo{Proof} By Theorem 4.2, any automorphism of Cend$_N$ has the form $\varphi_A(a(\p,x))=A(\p+x) a(\p, x+\alpha)A(x)^{-1}$, with $A(x)$ an invertible matrix in Mat$_N\CC[x]$. Suppose that  the restriction of $\varphi_A$ to $oc_{N,P}$ gives us an isomorphism between $oc_{N,P}$ and $oc_{N,Q}$. Then $\varphi_A(a(\p ,x)P(x))=A(\p+x) a(\p, x+\alpha)D(x)Q(x)$ for all $a(\p,x)\in oc_N$, where $D$ is an invertible matrix in Mat$_N\CC[x]$ and $P(x+\alpha)=D(x) Q(x)A(x)$. But the image is in $oc_{N,Q}$ if and only if (applying $\sigma_{Q,I,1,0}$)
$$
 a(\p,x-\alpha) R(x)=R^t(-\p-x) a(\p,x+\alpha) \quad\hbox{ for all } a(\p,x)\in oc_N,
 $$
where $R(x)=A^t(-x)D(x)^{-1}$. Therefore, we must have $\alpha=0$ and $R=c\,Id$  ($c\in\CC$), that is $D(x)=cA^t(-x)$. Hence $P(x)= cA^t(-x)Q(x)A(x)$, proving (a). Part (b) follows by similar arguments.
\qed\enddemo

A classification of finite irreducible subalgebras of $gc_N$ was given in [DK]. 
In view of the
 discussion of this section, it is natural to propose the following conjecture.
\vskip .3cm

\proclaim{Conjecture 6.20} Any infinite Lie conformal subalgebra of $gc_N$
acting irreducibly on $\CC [\partial]^N$ 
is conjugate by an automorphism of Cend$_N$ to one of the following subalgebras:

(a) $gc_{N,P}$, where $\det P\neq 0$,

(b) $oc_{N,P}$,  where $\det P\neq 0$ and $P(-x)=P^t(x)$,

(c) $spc_{N,P}$, where $\det P\neq 0$ and $P(-x)=-P^t(x)$.
\endproclaim

\vskip .4cm

\noindent{\bf Acknowledgment.} C. Boyallian and J. Liberati were supported in
 part by
Conicet, ANPCyT, Agencia Cba Ciencia,  
Secyt-UNC and Fomec (Argentina). V. Kac was supported in part by the NSF grant
DMS-9970007. Special thanks go to MSRI (Berkeley) for the hospitality during our stay there.

\enddocument